\documentclass[10pt,aps,prb,twocolumn,superscriptaddress,floatfix,english,twocolumn,amsmath,amssymmb,longbibliography]{revtex4-2}
\usepackage{amsfonts}
\usepackage{babel}
\usepackage{bm}
\usepackage{color}
\usepackage{comment}
\usepackage{dcolumn}
\usepackage{enumerate}
\usepackage{epsf}
\usepackage{subfigure}
\usepackage{epsfig,psfrag}
\usepackage{esint}
\usepackage{float}
\usepackage[T1]{fontenc}
\usepackage{graphicx}
 
\usepackage{latexsym}
\usepackage{relsize}
\usepackage[final]{hyperref} % adds hyper links inside the generated pdf file
\hypersetup{
	colorlinks=true,       % false: boxed links; true: colored links
	linkcolor=blue,        % color of internal links
	citecolor=blue,        % color of links to bibliography
	filecolor=magenta,     % color of file links
	urlcolor=blue         
}

\newcommand{\new}[1]{\textcolor{black}{#1}}

\newcommand{\ba}{\begin{array}}
\newcommand{\ea}{\end{array}}
\newcommand{\be}{\begin{equation}}
\newcommand{\ee}{\end{equation}}
\newcommand{\bea}{\begin{eqnarray}}
\newcommand{\eea}{\end{eqnarray}}

\begin{document}

\title{{Magnetic Bloch bands and Weiss oscillations in  Dirac mass superlattices}}

\author{Aman Anand}
\email{Aman.Anand@citystgeorges.ac.uk}
\affiliation{Department of Mathematics, City St George's, University of London, London EC1V 0HB, United Kingdom}

\author{Reinhold Egger} 
\email{egger@hhu.de}
\affiliation{Institut f\"ur Theoretische Physik, Heinrich-Heine-Universit\"at, Universit\"atsstr.~1, D-40225  D\"usseldorf, Germany}

\author{Alessandro De Martino}
\email{Alessandro.De-Martino.1@citystgeorges.ac.uk}
\affiliation{Department of Mathematics, City St George's, University of London, London EC1V 0HB, United Kingdom}	

\begin{abstract}
We study two-dimensional Dirac fermions in a one-dimensional mass superlattice 
under a perpendicular magnetic field. Using exact solutions for isolated and finite arrays of domain walls,
we demonstrate the persistence of Jackiw–Rebbi modes with a field-dependent renormalized velocity. For the periodic case, 
we adopt a gauge-invariant projection method onto magnetic Bloch states, valid for arbitrary fields and mass profiles, 
which yields dispersive Landau levels, and confirm its accuracy by comparison with finite arrays spectra. 
From the miniband spectra we predict modified quantum Hall plateaus and 
Weiss-like magnetoconductivity oscillations, characterized by a strongly reduced amplitude and a $\pi/2$ 
phase shift compared to electrostatic superlattices.
\end{abstract}

\maketitle

\section{Introduction}
\label{sec1}

Spatially periodic modulations of key parameters in quantum materials can provide a versatile route to miniband engineering and transport control. 
For example, by breaking translational symmetry on scales exceeding the underlying atomic crystal lattice constant, one can engineer periodic superlattices \cite{andrei2021,Song2015} which in turn give rise 
to a reconstruction of the band structure (in particular, the formation of minibands), unconventional quantum Hall sequences, and/or novel types of interface states. These periodic structures not only provide a means to control 
electronic properties but also serve as a fertile ground for exploring fundamental physics, including topological states and commensurability effects.

In this work, we study two-dimensional (2D) Dirac fermions subject to a 
constant perpendicular magnetic field $B$ in the presence of a  one-dimensional (1D) mass superlattice, where the Dirac mass $m(x)$ periodically alternates in sign along the (say) $x$-direction but remains independent of the $y$-coordinate. For $B=0$, this problem has been studied in Ref.~\cite{demartino2023}. 
Such types of mass superlattices acting on 2D Dirac fermions may be realized experimentally, for example, 
in graphene monolayers \cite{GrapheneReview2009,Song2015}, e.g., 
by depositing the layer onto a suitably patterned substrate. If the substrate potential oscillates 
on the scale of graphene's lattice constant, the potentials experienced by the two sublattices 
of the honeycomb lattice will differ and one effectively generates a mass term. 
Different types of superlattices (in particular, electrostatic superlattices) in graphene \new{(both monolayer and multilayer) and other 2D materials} have been studied theoretically 
\cite{tahir2007,Park2008a,Park2008b,Park2008c,Louie2009,Fertig2009,Nori2009,tahir2011,
Choi2014,Barbier2010,Barbier2010b,DellAnna2011,Lenz2011,Fertig2011,Chen2020,zarenia2012,maksimova2012,GhorashiPRL2023,GhorashiPRB2023,Song2023,Zeng2024,Seleznev2024,martelo2024,Tan2024,paul2025} 
and experimentally \cite{Deshmukh2013,Li2017a,Forsythe2018,Barrier2020,Ruiz2022,Dean2021,Hong2024,Sun2024}.
Similarly, 2D Dirac fermions can be realized as surface states in three-dimensional topological insulator slabs \cite{Hasan2010}, where the mass superlattice could be generated by decorating the surface with ferromagnetic strips of alternating magnetization direction. 

If the mass modulation $m(x)$ corresponds to a finite sequence of constant mass regions with alternating sign, i.e., 
a finite-size array structure, one can obtain the exact spectra by means of the transfer matrix method.
In particular, we show that the well known chiral Jackiw–Rebbi  
interface mode localized at a single mass kink for $B=0$ \cite{JackiwRebbi1976,Jackiw1981,Semenoff2008} 
will persist in a finite magnetic field.  
However, as a consequence of the $B$ field, the mode velocity is renormalized to smaller values when increasing $B$. 
We show that this effect is robust against spatial variations of the precise mass kink profile, i.e., 
the effect occurs both for sharp and smooth mass domain walls.
We also compute the band structure for finite-size array structures with $N$ kink-antikink configurations 
by means of the transfer matrix approach. However, for very large $N$, this approach becomes impractical.

For the infinite mass superlattice case, we instead adopt the gauge-invariant 
projection method onto magnetic Bloch states (MBSs) developed in Ref.~\cite{Bernevig2022}. 
Remarkably, in contrast to 2D superlattices where commensurability constraints apply,
in the 1D case this approach is applicable for arbitrary magnetic fields 
and arbitrary periodic mass profiles $m(x)$.  By comparing the corresponding band structure results to the 
transfer matrix results for finite-size array structures with $N=7$, 
we find already good agreement between both approaches, thus providing a valuable consistency check.
The MBS projection approach then allows for an efficient computation of the magnetic miniband spectrum in the mass superlattice case. 
We also formulate a complementary low-energy theory describing the strong magnetic field regime.

Using these theoretical tools, we then analyze charge transport properties in this system. In particular, 
we show how the mass superlattice modifies the relativistic (half-integer) quantum Hall plateaux \cite{Gusynin2005}.  
Moreover, we investigate if and how Weiss oscillations of the longitudinal magnetoconductivity $\sigma_{yy}$ 
as a function of $1/B$ are possible in a mass superlattice. We recall that for a conventional 2D electron gas 
subject to a 1D electrostatic superlattice and a constant magnetic field, commensurability effects give rise to
Weiss oscillations \cite{Weiss1989,Winkler1989,Gerhardts1989,Beenakker1989,Pfannkuche1992,Peeters1992}.  
Such effects have already been observed in graphene \cite{Eroms2018,Dean2021,Huber2022,Paul2022} 
and in topological insulators \cite{Koop2024}.  Compared to the corresponding electrostatic superlattice case 
for 2D Dirac fermions \cite{Matulis2007,tahir2007}, we find that Weiss oscillations are also present in 
a mass superlattice.  However, we predict a significant suppression of the oscillation amplitude
together with a characteristic $\pi/2$ phase shift relative to the electrostatic case.

The remainder of this paper is organized as follows. In Sec.~\ref{sec2}, we introduce the model and describe 
its solution for a single mass kink profile and for an array of mass kinks and antikinks by means of the transfer matrix technique. 
In Sec.~\ref{sec3}, we construct the magnetic Bloch bands (MBBs) in mass superlattices by developing and applying the MBS projection approach. 
Charge transport observables are then studied in Sec.~\ref{sec4}. 
Finally, Sec.~\ref{sec5}  offers some concluding remarks. Technical details are collected in several Appendices. 

\section{Model and exact solution}
\label{sec2}

We consider 2D Dirac fermions subjected to a 1D mass superlattice described by a spatially varying profile $m(x)$, which  
alternates between positive and negative values with period $d$ \cite{zarenia2012,demartino2023}. 
In the presence of a uniform perpendicular magnetic field $B>0$, 
the system is described by the Hamiltonian
\begin{align}
\mathcal{H}  & = v_\mathrm{F} \bm{\sigma}  \cdot \bm{\Pi}  + m(x)\sigma_z+V(x),
\label{ham}
\end{align}
where  $\mathbf{\Pi} =  -i\hbar \nabla + e \mathbf{A}$ is the \new{kinematic} momentum, $v_{\text F}$ is the Fermi velocity,
and $\bm{\sigma}=(\sigma_x,\sigma_y)$ and  $\sigma_z$ are the standard Pauli matrices. We focus on the case of a constant electrostatic potential $V$, which merely shifts the energy and will henceforth be omitted. 

The model \eqref{ham} encapsulates the low-energy physics of electronic states in 2D graphene monolayers close to 
a single Dirac node (valley) and for a definite spin polarization. This approximation is justified if the mass modulation 
is smooth on the scale of graphene's lattice constant \cite{GrapheneReview2009} and if the (typically small) Zeeman term is neglected. 
A periodic mass term may arise from a suitably patterned substrate that creates a sublattice-symmetry breaking term with 
domain walls between regions with different signs of the staggered on-site energy \cite{Semenoff2008}. 
By means of proximity screening, recent experimental progress \cite{Geim2025} has shown that disorder effects can be largely eliminated, 
resulting in ultraclean graphene samples.  We thus neglect disorder effects in what follows with the exception of Sec.~\ref{sec4b}. 
Electron-electron interaction effects are suppressed in a natural way by proximity screening.  
We therefore also neglect interactions below.  In addition to graphene monolayers,  
the model \eqref{ham} describes the spin-momentum locked and protected 
surface states in three-dimensional topological insulators \cite{Hasan2010,Shen2017}. 
In that case, the periodic mass term could be engineered by depositing ferromagnetic stripes with alternating magnetization on the material surface.  

In this section, we work in the Landau gauge $\mathbf{A}=(0,Bx)$. Then translation invariance in the $y$-direction 
implies that the wave number $k_y$ is conserved and thus  the wave function can be expressed in the form
\begin{equation} 
  \Psi(x,y) = \frac{e^{ik_yy}}{\sqrt{L_y}} \psi(x),  
  \label{2Dwf}
\end{equation}
where $L_y$ is a normalization length and we omit the dependence of $\psi(x)$ on $k_y$ for notational simplicity. 
The Dirac equation reduces to the 1D problem $\mathcal H (k_y)\psi(x) = E\psi(x)$,
with the Hamiltonian
\begin{align}
\mathcal H (k_y)&=v_\mathrm{F}\sigma_x\hat p_x+v_\mathrm{F}\sigma_y (\hbar k_y + eBx) + m(x)\sigma_z ,
\label{hamMF}
\end{align}
where $\hat p_x=-i\hbar\partial_x$. In the remainder of this section, 
we present the solution of this model for particular mass profiles before turning to the periodic mass superlattice in Sec.~\ref{sec3}.

We define the cyclotron energy $\varepsilon_c$ and the magnetic length $\ell_B$ as 
\begin{equation}\label{scalesdefinition}
\varepsilon_c= v_F\sqrt{2\hbar eB}, \quad \ell_B=\sqrt{\frac{\hbar}{eB}},
\end{equation}
and use them as units of energy and length, respectively. When analyzing 
the dependence on magnetic field, we express all quantities in terms of the cyclotron energy and the magnetic length at $B=B_0=1$\,Tesla, 
denoted by $\bar \varepsilon_c$ and $\bar \ell_B$. With $v_{\text F}= 10^6$\,m/s \cite{GrapheneReview2009}, 
these quantities evaluate to 
\begin{equation}\label{scalesdef2}
\bar \varepsilon_c \simeq 36\,{\rm meV},\quad \bar \ell_B\simeq 26\,{\rm nm}.
\end{equation} 

The rest of this section is organized as follows.  
In Sec.~\ref{sec2a}, we provide the general solution in a region of constant mass. In Sec.~\ref{sec2b}, 
we discuss the solution for a single mass kink (domain wall), followed by the case of a finite array of kinks 
and antikinks in Sec.~\ref{sec2c}. Finally,  in Sec.~\ref{sec2d}, we outline a general approach valid for slowly varying mass profiles 
and strong magnetic fields.
Throughout the paper, we set $\hbar=v_\mathrm{F}=1$ unless stated otherwise.

\subsection{Uniform mass}
\label{sec2a}

The general eigenfunction of the Hamiltonian~\eqref{hamMF}
with constant mass $m(x)=M$ can be cast in the form \cite{demartino2023}
\begin{equation}
\psi(x) = 
 W_{M}(x) \begin{pmatrix} 
a \\ b
\end{pmatrix},
\end{equation}
where $a$ and $b$ are arbitrary complex coefficients.
The matrix $W_M(x)$ is given by 
\begin{equation}
    W_M(x) =     \begin{pmatrix}    \frac{E+M}{\varepsilon_c} D_{p-1}(-q) & \frac{E+M}{\varepsilon_c} D_{p-1}(q)  &  \\
     -iD_p(-q) & iD_p(q)  \end{pmatrix},
\label{WmatrixApp}
\end{equation}
where $D_p(q)$ 
is the parabolic cylinder function \cite{NIST:DLMF} with the definitions
\begin{equation}\label{pqxdef}
    p=\frac{E^2-M^2}{\varepsilon_c^2},\quad
 q= \frac{\sqrt{2} ( x-x_c )}{ \ell_B}, \quad x_c=-k_y\ell_B^2. 
\end{equation}

For a system of infinite extent, the requirement of normalizability implies that $p$ in Eq.~\eqref{pqxdef}
can only take nonnegative values, which leads to the well-known Landau levels \cite{GrapheneReview2009}  
\begin{align}
E_0 = -M, \quad E_{n}  = s_n  \sqrt{\varepsilon_c^2|n|+M^2}, \quad n\in \mathbb{Z}^*,
\label{DLLfiniteM}
\end{align}
where we define $s_n=\mathrm{sgn}(n)$ with $s_0=0$, and $\mathbb{Z}^*$ the set of nonzero integers.
The 2D normalized eigenstates for $M=0$, which will be used
in Sec.~\ref{sec4b} below, read
\begin{equation}
     \Psi_{n,k_y}(x,y) = \frac{\mathcal{N}_{n}e^{ik_y y}}{\sqrt{L_y \ell_B}} 
     \begin{pmatrix}
        s_{n} \Phi_{|n|-1} (\frac{x-x_c}{\ell_B} ) \\
        i\Phi_{|n|} (\frac{x-x_c}{\ell_B})
    \end{pmatrix}, 
    \label{wf}
\end{equation}
where $\mathcal{N}_n=(2-\delta_{n,0})^{-\frac{1}{2}}$. 
The harmonic oscillator eigenfunctions $\Phi_n(x)$ ($n\geq 0$) 
can be expressed in terms of Hermite polynomials \cite{NIST:DLMF},
\begin{equation}
    \Phi_n (x)  = \frac{1}{\sqrt{n!\sqrt{\pi}}} D_n(\sqrt{2}x)    
    = \frac{e^{-x^2 /2}}{\sqrt{2^n n! \sqrt{\pi}}} H_n (x).
\end{equation}
 The wave functions~\eqref{wf} will be used in the perturbative approach of Sec.~\ref{sec4b}.

\subsection{Single mass kink}
\label{sec2b}

Next, we study a mass profile featuring a single domain wall. 
Specifically, we consider a sharp kink of the form
\begin{equation}
m(x)=M\text{sgn}(x), \quad M>0. 
\label{kink}
\end{equation}
Since this profile lacks an intrinsic length scale, 
the spectrum depends only on the dimensionless parameter $M/\varepsilon_c$. 
In App.~\ref{AppA}, we present the solution for a smooth kink profile and show that the spectral properties 
are only weakly affected by the smoothness parameter. 
In the absence of a magnetic field, a domain wall as defined by Eq.~\eqref{kink} is well known 
to host a topologically protected 1D chiral mode, the so-called Jackiw-Rebbi mode, which propagates unidirectionally along the $y$-direction \cite{JackiwRebbi1976,Semenoff2008,Zutic2021,Wang2021a}.
As we show below, this mode survives the presence of a magnetic field but its group velocity becomes suppressed. 

Taking into account the requirement of normalizability, 
the wave function can be written as 
\be
\psi(x) = \left\{
\begin{array}{ll}
  W_{-M}(x) \begin{pmatrix} 
a_L \\ 0
\end{pmatrix}   &  \text{for} \; x<0 \\
W_{M}(x) \begin{pmatrix} 
0 \\ b_R
\end{pmatrix}    &  \text{for} \;  x> 0 
\end{array} \right.,
\ee
where $a_L$ and $b_R$ are complex coefficients.
Continuity of the wave function at the kink position $x=0$ 
requires
\be
\begin{pmatrix} 
0 \\ b_R
\end{pmatrix} = \bm{ \bm{\Omega}}(0)
\begin{pmatrix} 
a_L \\ 0
\end{pmatrix},
\label{quant1}
\ee
where we define the transfer matrix
\begin{align}
\bm{\bm{\Omega}}(x) &= W^{-1}_{ M}(x) \, W_{- M}(x)  .
\label{omega}
\end{align}
Equation~\eqref{quant1} relates the coefficient $b_R$ to $a_L$, which is
then fixed by the overall normalization. A nontrivial solution exists only if 
the condition $\Omega_{11}(0)=0$ is satisfied. This equation 
yields the explicit quantization condition
\be
\frac{(E-M)D_p(\sqrt{2}k_y\ell_B) D_{p-1}(-\sqrt{2}k_y\ell_B)}{(E+M)D_{p-1}(\sqrt{2}k_y\ell_B)D_{p}(-\sqrt{2}k_y\ell_B)}= -1,
\label{QCkinkMF}
\ee
with $p$ in Eq.~\eqref{pqxdef}.
Note the invariance of Eq.~\eqref{QCkinkMF} under the transformation $(k_y,E)\rightarrow (-k_y,-E)$. 

\begin{figure}[t]
  \centering
  \subfigure[]{\includegraphics[width=0.48\linewidth]{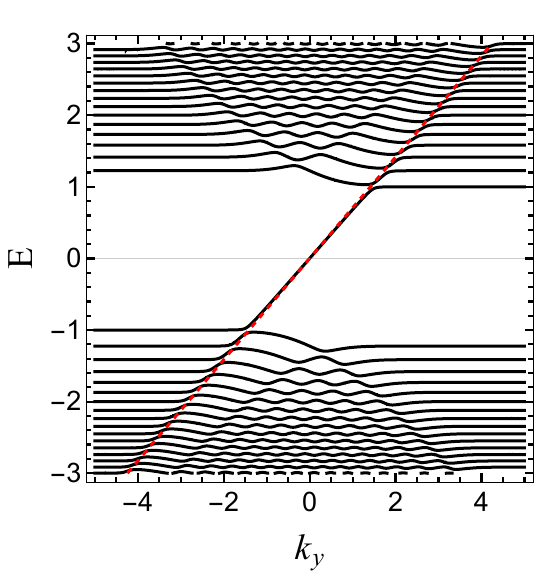}}
  \subfigure[]{\includegraphics[width=0.48\linewidth]{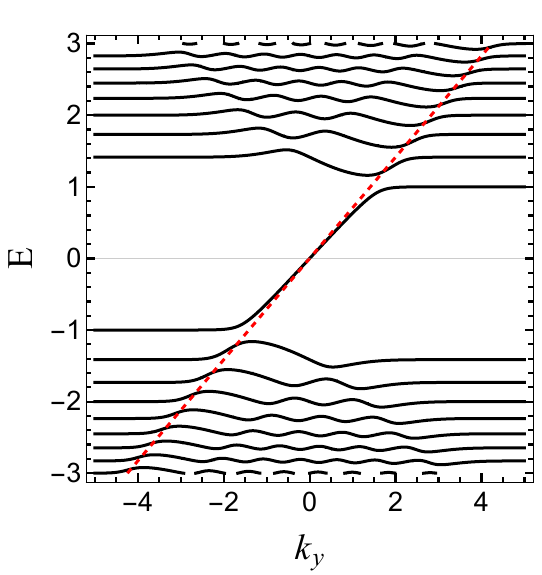}} \\
  \subfigure[]{\includegraphics[width=0.49\linewidth]{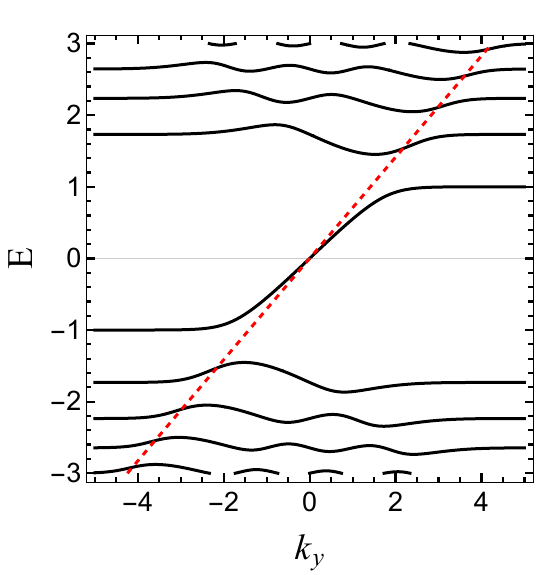}}
  \subfigure[]{\includegraphics[width=0.49\linewidth]{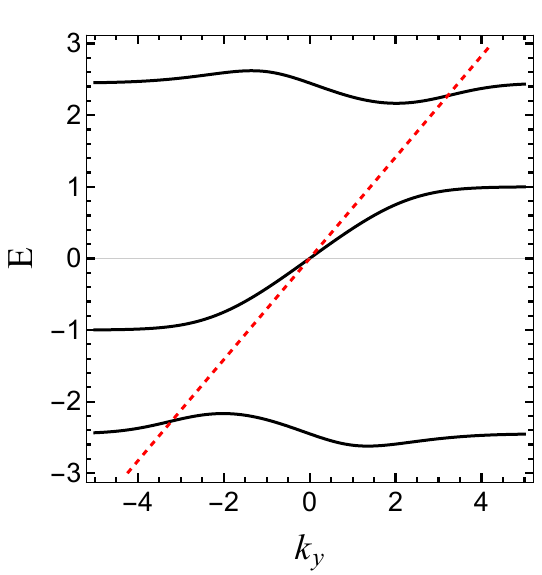}}
  \caption{Evolution of the energy spectrum for the mass profile \eqref{kink} with a single mass kink, 
  obtained by numerical solution of Eq.~\eqref{QCkinkMF} with $M=\bar \varepsilon_c$, 
  as the magnetic field increases: $B=0.5, 1, 2, 5$~T, see panels (a), (b), (c), and (d), respectively.
Energy is expressed in units of $\bar \varepsilon_c$ and $k_y$ in units of $\bar \ell_B^{-1}$, see Eq.~\eqref{scalesdef2}.
For comparison, we also plot the linear dispersion of the interface chiral mode at $B=0$ (red dashed line). }  
\label{fig1}
\end{figure}

Numerical solution of Eq.~\eqref{QCkinkMF} leads to the spectrum shown in Fig.~\ref{fig1}.
We observe that due to the kink in the mass profile, the Landau levels acquire dispersion.  
For $M \ll \varepsilon_c$, the mass kink does not play a significant  role, 
and we find almost perfectly flat Landau levels. For larger and larger $M$, Landau levels 
acquire a stronger dispersion, especially close to $k_y=0$. For $M\gg\varepsilon_c$, one recovers 
the usual chiral interface mode with linear dispersion which exists for $B=0$. 
However, a finite magnetic field renormalizes the velocity of the interface chiral Jackiw-Rebbi mode, $v_{\rm F}\to v_r$, 
as we discuss next. 

\begin{figure}[ht!]
\includegraphics[width=.8\linewidth]{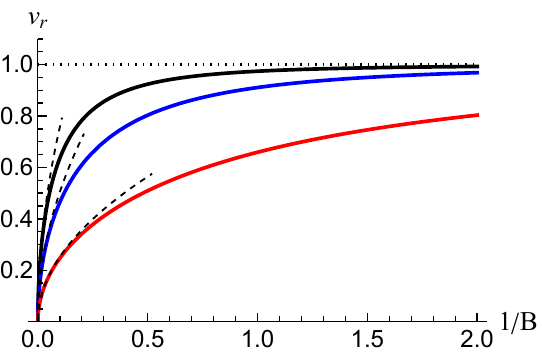}
\caption{Renormalized velocity $v_r$ (in units of $v_{\text F}$)
of the chiral 1D Jackiw-Rebbi mode propagating along a single sharp mass kink, see Eq.~\eqref{renvelocity}, 
vs inverse magnetic field $1/B$ (with $B$ given in Tesla) 
for different mass amplitudes: $M=1.5\bar \varepsilon_c$ (black),  
$M=\bar \varepsilon_c$ (blue), $M=0.5\bar \varepsilon_c$ (red),  
with $\bar \varepsilon_c$ in Eq.~\eqref{scalesdef2}. The black dashed curves 
illustrate the asymptotic $B^{-\frac{1}{2}}$ scaling at large field.
The black dotted line shows the Jackiw-Rebbi mode velocity at $B=0$.}
\label{fig2}
\end{figure}

By expanding the quantization condition \eqref{QCkinkMF} around $(E,k_y)=(0,0)$, one can find 
an analytical expression for the magnetic-field dependent velocity of the chiral Jackiw-Rebbi mode, $v_r(M/\varepsilon_c)$, 
see Eq.~\eqref{scalesdefinition} for the definition of $\varepsilon_c(B)$. With $\xi=M/\varepsilon_c$, we obtain 
\begin{align}
v_r(\xi) & = 2^{3/2 -\xi^2}  \sqrt{\pi} \xi \,  \Gamma(1 + \xi^2) \times \nonumber \\
&\times \left(\frac{1}{\Gamma^2( 1/2 + \xi^2/2)}-
\frac{1}{\Gamma(\xi^2/2) \, \Gamma(1 + \xi^2/2)} 
\right), 
\label{renvelocity}
\end{align}
where $\Gamma(x)$ denotes the Gamma function \cite{NIST:DLMF}.
For $M/\varepsilon_c\ll 1$, corresponding to the large-field limit, the renormalized velocity becomes very small 
and scales with magnetic field as $v_r\sim \sqrt{\frac{8}{\pi}}\frac{M}{\varepsilon_c}\propto B^{-\frac{1}{2}}$.  
On the other hand, for $B\to 0$, corresponding to $M/\varepsilon_c\gg 1$, one finds that $v_r$ approaches $v_{\rm F}$. 
The full dependence of $v_r$ on the (inverse) magnetic field based
on Eq.~\eqref{renvelocity} is depicted in Fig.~\ref{fig2}.

\subsection{Finite array of alternating domain walls}
\label{sec2c}

\begin{figure}[t!]
\includegraphics[width=9cm]{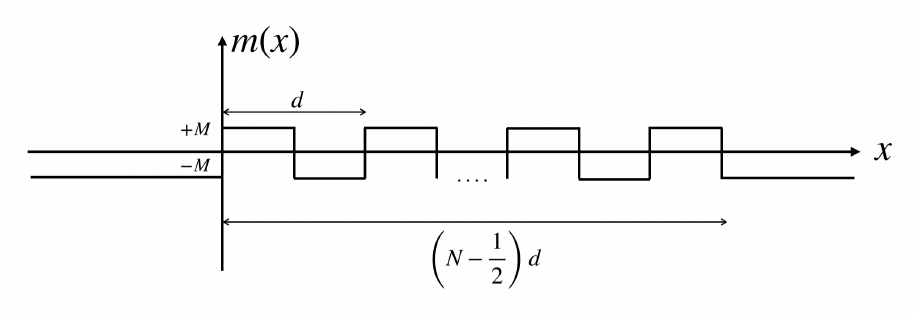}
\caption{Mass profile $m(x)$ for a finite array of $N$ kinks and antikinks, with width $L=(N-\frac{1}{2})d$.}
\label{fig3}
\end{figure}

Next, we consider an array of $N$ mass kinks and antikinks, with inter-kink spacing $d$, 
assuming that the mass is constant outside a finite interval,   
see Fig.~\ref{fig3},
\begin{equation}\label{finitearrayprofile}
        m(x) =  \left\{
        \begin{array}{cc}
        - M & \quad  x<0 \quad \text{and} \quad  x > Nd,  \\
        + M &  \quad 0 < x- jd < \frac{d}{2}, \\
        - M &  \quad  \frac{d}{2} < x -jd < d, 
    \end{array} \right. 
\end{equation}
with $j=0,1,2,\dots,N-1$. 
The wave function can be written as
 \begin{equation}
 \label{Eqn:finite_array_wavefn}
        \psi(x)=  \left\{
        \begin{array}{lc}
        W_{-M}(x)
\begin{pmatrix} 
a_0 \\ b_0
\end{pmatrix} & \quad x < 0 \\
        W_{+M}(x)
\begin{pmatrix} 
a_{2j+1} \\ b_{2j+1}
\end{pmatrix} & \quad  0 < x - jd <\frac{d}{2} \\       
        W_{-M}(x)
\begin{pmatrix} 
a_{2j+2} \\ b_{2j+2}
\end{pmatrix} 
     &  \quad \frac{d}{2} < x -jd  < d \\
        W_{-M}(x)
\begin{pmatrix} 
a_{2N} \\ b_{2N}
\end{pmatrix} & \quad  Nd < x
    \end{array} \right. .
\end{equation}
To ensure that it is normalizable, we impose the boundary conditions
\begin{equation}
    \begin{pmatrix} 
a_0 \\ b_0
\end{pmatrix}  = \begin{pmatrix} 
a_L \\ 0
\end{pmatrix} , \quad \begin{pmatrix} 
a_{2N} \\ b_{2N}
\end{pmatrix}  = \begin{pmatrix} 
0 \\ b_R
\end{pmatrix} ,
\end{equation}
where $b_R$ is related to $a_L$ by the continuity requirement, see Eq.~\eqref{leftrightcoeff} below,
and $a_L$ is then fixed by normalization. 
The continuity conditions at the position of the kinks and antikinks read
\begin{align}
   W_{-M}(jd)
     \begin{pmatrix} 
a_{2j} \\ b_{2j}
\end{pmatrix}  & = W_{+M}(jd)\begin{pmatrix} 
a_{2j+1} \\ b_{2j+1}
\end{pmatrix} ,  \\
    W_{+M}(jd+\frac{d}{2})
     \begin{pmatrix} 
a_{2j+1} \\ b_{2j+1}
\end{pmatrix} & =  W_{-M}(jd+\frac{d}{2})
\begin{pmatrix} 
a_{2j+2} \\ b_{2j+2}
\end{pmatrix}. \nonumber
\end{align}

Using these equations recursively, 
we obtain a relation between the coefficients on the right and left sides, 
\begin{equation}
    \begin{pmatrix} 
0 \\ b_R
\end{pmatrix} = \bm{\Omega}^{(N)}
\begin{pmatrix} 
a_L \\ 0
\end{pmatrix}.
\label{leftrightcoeff}
\end{equation}
The transfer matrix for the array, $\bm{\Omega}^{(N)}$,  reads
\begin{align}
    \bm{\Omega}^{(N)}  = \bm{\Omega}^{-1} (Nd-\frac{d}{2}) \, 
    \bm{\Omega} (Nd-d)  \cdots  \bm{\Omega}^{-1}(\frac{d}{2}) \,  \bm{\Omega}(0) , 
\end{align}
with $\bm{\Omega}(x)$ in Eq.~\eqref{omega}. The quantization condition is then given by
\begin{equation}
    \Omega^{(N)}_{11} = 0.
    \label{Eqn:finite_array1}
\end{equation}

For symmetry reasons, it is convenient to move the center of the array to the origin. 
This amounts to replacing $m(x)\to \widetilde{m}(x)=m(x-L/2)$, where $L=Nd-\frac{d}{2}$ is the array width.  
The relation $\widetilde{m}(-x)=\widetilde{m}(x)$ then implies that the Hamiltonian $\mathcal{H}(x,k_y)$ enjoys inversion symmetry,
\begin{equation}
    \sigma_z \mathcal{H}(x,k_y) \sigma_z= \mathcal{H}(-x,-k_y),
    \label{invsymm}
\end{equation}
which means that the spectrum is symmetric under $k_y\rightarrow -k_y$, i.e., $E(-k_y)=E(k_y)$.
Moreover, the quantity $x_c=-k_y\ell_B^2$ in Eq.~\eqref{pqxdef} corresponds to the position of the guiding center 
relative to the center of the array. Since the wave functions are localized on the scale of the
magnetic length $\ell_B$, they are insensitive to the array boundaries as long as both $\ell_B\ll L$ and $|k_y\ell_B^2| \ll \frac{L}{2}$.
As a result, for sufficiently large $N$ and strong magnetic field, the energy spectrum of the finite array
closely approximates that of the periodic system. Indeed, when analyzing the periodic case,  
we find excellent agreement, see Fig.~\ref{fig6} below. 

\begin{figure}[t]
  \centering
  \includegraphics[width=\linewidth]{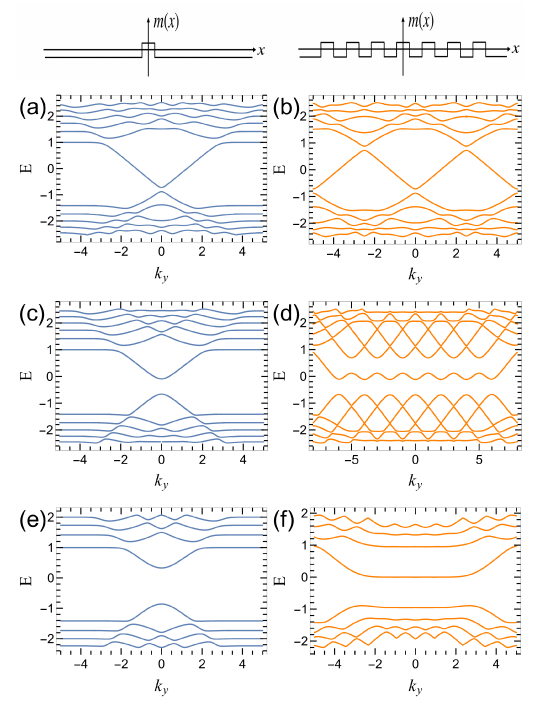}
  \caption{Energy spectra for finite-length kink-antikink arrays with the mass profile in Eq.~\eqref{finitearrayprofile} for $M=\bar\varepsilon_c$.
  We set $B=1$\,T and use $\bar \varepsilon_c$ and $\bar \ell_B^{-1}$ in Eq.~\eqref{scalesdef2}   as units for energy and $k_y$, respectively.
  \new{The mass profiles $m(x)$ for the left and right columns are indicated schematically on top of the figure. 
  Results are shown for $N=1$ (indigo) and $N=7$ (orange curves) kink-antikink pairs and different values of the inter-kink spacing $d$.
  Panels (a) and (b) are for $d=5\bar\ell_B$. Panels (c) and (d) are for $d=2\bar\ell_B$.  Panels (e) and (f) are for $d=\bar\ell_B$.}
   }
\label{fig4}
\end{figure}

\begin{figure}[t]
\includegraphics[width=.9\linewidth]{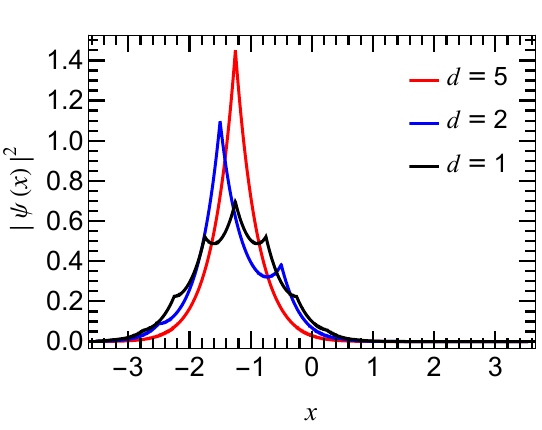}
\caption{\new{Probability density profile for the state in the "zero-energy" band at $k_y =1.25\bar\ell_B ^{-1}$ for $B=1$\,T, $N = 7$ and different $d$ values, see Figs.~\ref{fig4}(b,d,e).  The corresponding energies are $E =0$, $0.081784\bar\varepsilon_c$, and $0.000221\bar\varepsilon_c$, for $d/\ell_B = 5,2,1$, respectively. }}
\label{fig5}
\end{figure}

Numerical solution of Eq.~\eqref{Eqn:finite_array1} results in the spectra shown in Fig.~\ref{fig4} for two different array sizes $N$ 
and various values of the inter-kink spacing $d$. As expected from the inversion symmetry \eqref{invsymm}, the spectra 
are symmetric under $k_y\rightarrow -k_y$.
\new{In Figs.~\ref{fig4}(a,b), corresponding to the largest separation $d$, the kinks are far away from each other.} 
As a result, a wave function centered at $x_c=-k_y\ell^2_B$ 
\new{has only a small overlap with both the nearest kink and antikink configurations.
Indeed, once the separation between kinks and antikinks 
exceeds the magnetic length, energy levels are dispersive with only small band gaps. In this regime, Jackiw-Rebbi states 
are clearly visible as linearly dispersive modes near zero energy.  
For an array with $N$ kink-antikink pairs, there are $N$ right-moving and $N$ left-moving Jackiw-Rebbi modes. In addition,
for $N=7$, the band closest to zero energy exhibits oscillations in $k_y$
with period $d/\ell_B^2$. The amplitude of these energy band oscillations is of order $\sim M$, reflecting the underlying mass modulation. 
Next, Figs.~\ref{fig4}(c,d) and Figs.~\ref{fig4}(e,f) show numerical results for smaller values of $d$. 
In that case, the wave function at a given $k_y$ overlaps both the nearest kink and antikink configurations. 
Since the effects of kinks and antikinks then tend to average out, one arrives at rather flat energy bands separated by large gaps. 
This feature is especially pronounced for $N=7$, where nearly flat bands appear close to zero energy 
over a range of $k_y$ corresponding to guiding centers located inside the array. 
Within this region, the energy levels exhibit oscillations as function of $k_y$ with period $d/\ell_B^2$. 
While the oscillation amplitude is too small to be visible on the shown scales in Fig.~\ref{fig4}(f), 
they are clearly observable, e.g., in Fig.~\ref{fig4}(d). For small $d$, we conclude that the putative 
Jackiw-Rebbi states localized near individual kinks or antikinks are strongly hybridized 
and do not fully develop into distinct modes. }

\new{This behavior is illustrated in Fig.~\ref{fig5}, where, using Eq.~\eqref{Eqn:finite_array_wavefn}, 
we show how the real-space probability density profile of a state in the "zero-energy" band at given $k_y$ evolves with $d$. 
The profile for $d=5\bar\ell_B$ features a single, relatively narrow peak, corresponding 
to a Jackiw-Rebbi state localized at a kink. As $d$ decreases, the profile broadens and develops multiple peaks, 
illustrating that the state arises from the hybridization of several adjacent Jackiw-Rebbi states.}

It is worth pointing out that we have also studied other related 
mass profiles for finite-size arrays. One variant corresponds to the configuration shown in Fig.~\ref{fig3}, 
but with the mass vanishing outside the array.
Another starts with a kink at $x=0$ and ends with a kink at  $x=Nd-d$, resulting in
a mass term with opposite sign at $x\rightarrow \pm \infty$.
In both cases, the energy spectra were found qualitatively similar to those discussed above. 

\subsection{Strong magnetic fields}
\label{sec2d}

In the limit of strong magnetic fields, one can derive an approximate 
analytical expression for the dispersive Landau levels associated with the Hamiltonian in Eq.~\eqref{hamMF},
valid for an arbitrary smooth mass profile $m(x)$. This approximation is valid as long as 
the characteristic length scale over which $m(x)$ changes is much larger than the magnetic length $\ell_B$.
Additionally, we assume that the cyclotron energy $\varepsilon_c$ is much larger than the typical amplitude $M$ of the mass modulation. 

In this regime, wave functions are strongly localized at the guiding center, $x_c=-k_y\ell_B^2$,
and are sensitive only to the local value $m(x_c)$ and to the slope $m'(x_c)$ of the mass profile at that point. 
Effectively, the problem then reduces to one with a linear mass profile, with both the constant term and the slope determined 
by $k_y$ since $x_c$ depends on $k_y$. Under the above conditions, this approach provides a good approximation 
to the energy spectrum and, for a periodic mass profile, to the low-energy band structure across the entire 1D Brillouin zone.

Anticipating that the wave functions are strongly localized at $x=x_c$, we make the change of variable 
$x \rightarrow \tilde x = x-x_c$ and expand the mass term around $\tilde x=0$ to linear order.
We thus approximate $\mathcal H (k_y) \approx \mathcal H_\text{lin} (k_y)$ with
\begin{align}
\mathcal H_\text{lin} (k_y) 
  & = \sigma_x \hat p_{\tilde x}  +  eB \tilde x \sigma_y + \left[ m(x_c) + m'(x_c)  \tilde x\right] \sigma_z,
\end{align}
where $m'(x_c)=\partial_xm|_{x_c}$.
Using the results of App.~\ref{AppB}, we obtain the dispersive Landau levels 
 \begin{align}
 E_0(k_y) &= -m(x_c)   \cos \alpha ,  \\ \nonumber
    E_{n}(k_y)  &= s_n \sqrt{ 2 e \widetilde B|n|+ m^2(x_c)   \cos^2 \alpha },  \quad n\in\mathbb{Z}^*,
\end{align}
where we define  
\begin{equation}
    \widetilde B(k_y) = \sqrt{B^2+(m'(x_c)/e)^2}, \quad \cos\alpha = \frac{B}{\widetilde B(k_y)}.
\end{equation}
We recall that $s_n={\rm sgn}(n)$ with $s_0=0$ and $\mathbb{Z}^*$ is the set of nonzero integers, 
see the line below Eq.~\eqref{DLLfiniteM}. As expected, the energy levels
depend on the local value of the mass, while the local slope of the mass profile
renormalizes the value of the magnetic field $B\to \widetilde B$. For a periodic mass profile with period $d$,
the spectrum is explicitly periodic in $k_y$, with the period $d/\ell^2_B$ inherited from 
the mass superlattice. We note that this approximation is self-consistent, as the resulting eigenstates 
are localized on the scale $\widetilde \ell_B =1/\sqrt{e\widetilde B}$, which is shorter than $\ell_B$.
%i.e., we have $\widetilde\ell_B\le \ell_B$.

\section{Mass superlattice}
\label{sec3}

In this section, we turn to the case of a periodic mass term.
To make contact with the previous section, we focus on a profile featuring
sharp kinks and antikinks, given by 
\begin{equation}
m(x) =  \left\{
        \begin{array}{lc}
        +M, & \quad  | x - jd | < \frac{d}{4}, \\
        -M, &  \quad \frac{d}{4} \leq | x - jd | \leq \frac{d}{2},
    \end{array} \right. \quad j\in \mathbb{Z}.
    \label{periodicprofile}
\end{equation}
However, the approach presented below can be applied 
to an arbitrary 1D mass superlattice. 

The transfer matrix method described in Sec.~\ref{sec2}, which works 
very efficiently for the case of finite  arrays with $N$ kinks and antikinks, becomes 
impractical as $N\rightarrow  \infty$. Moreover, it relies on a specific gauge 
choice and on the existence of an exact solution, which is available 
only for very specific mass profiles. 
In what follows, we therefore adopt the alternative approach of Ref.~\cite{Bernevig2022}, 
which is conceptually transparent and proceeds in two steps, \new{briefly summarized below in order to keep the paper self-contained}. First, a basis of 
MBSs respecting the periodicity of the mass superlattice is constructed from
the Landau eigenstates obtained
in the absence of a mass term. Second, the full Hamiltonian is projected 
onto this basis, resulting in an infinite-dimensional matrix representation 
that can be truncated and diagonalized numerically. This approach allows one to calculate  
the magnetic band structure and the corresponding eigenstates to the desired level of accuracy, 
limited only by the available computational resources. This approach has the advantage 
of being explicitly gauge invariant and applicable to arbitrary periodic mass terms. 

In its original implementation~\cite{Bernevig2022}, the method was developed 
for twisted bilayer graphene, where massless Dirac electrons experience
a 2D moir\'e potential. In that geometry, the approach works only at
commensurate magnetic fields, i.e., when the magnetic flux 
through the superlattice unit cell in units of the elementary flux quantum $\Phi_0=h/e$, 
denoted as $\phi/2\pi$, is an integer. In particular,
the case $\phi=2\pi$ was examined in detail in Ref.~\cite{Bernevig2022}.
In contrast, no such restriction arises for our 1D mass superlattice, since
the system remains translationally invariant in the $y$-direction. One can therefore
introduce a fictitious periodicity along this direction such that, for any given magnetic
field $B$, the magnetic flux through the corresponding rectangular unit cell 
is exactly $\Phi_0$. 

Let us then consider the Hamiltonian ~\eqref{ham}, which we write as
$\mathcal{H}=\mathcal{H}_0+m(x) \sigma_z$. The mass term, viewed as a function 
of the 2D coordinate $\mathbf{r}$, is invariant under discrete translations 
by the superlattice vectors
\begin{equation}
    \mathbf{R} = R_1  \mathbf{a}_1 + R_2 \mathbf{a}_2, 
    \label{SuperlatticeVectors}
\end{equation}
where $R_i\in\mathbb{Z}$ and the primitive lattice vectors are given by $\mathbf{a}_1 = (d,0)$ and $\mathbf{a}_2 = (0,d_y)$.
For now, $d_y$ is an arbitrary but fixed period along the $y$-direction, and the area of the superlattice 
unit cell is $\mathcal{A} =|\mathbf{a}_1 \times \mathbf{a}_2 |= dd_y$. 
The reciprocal lattice is spanned 
by the vectors $2\pi  \mathbf{G}$ with
\begin{equation}
  \mathbf{G} = G_1  \mathbf{b}_1 + G_2 \mathbf{b}_2, 
\label{reciprocallatticevectors}
\end{equation}
where $G_i\in\mathbb{Z}$ and the basis vectors $\mathbf{b}_1 = (1/d, 0)$ and $\mathbf{b}_2 = (0,1/d_y)$ satisfy the orthogonality relation $\mathbf{a}_i \cdot \mathbf{b}_j = \delta_{ij}$. 
We formally recover a 2D superlattice problem analogous to the one studied 
in Ref.~\cite{Bernevig2022}, and hence we can apply their analysis to our system.

In the remainder of this section, we first recall the gauge-invariant 
formulation of Landau states, see Sec.~\ref{sec3a}. 
We then  construct the basis of MBSs without mass term in 
Sec.~\ref{sec3b}, and finally compute the MBBs by numerically diagonalizing the truncated Hamiltonian matrix in Sec.~\ref{sec3c}.

\subsection{Gauge-invariant formulation}
\label{sec3a}

We start by introducing the \new{kinematic} momenta $\Pi_\mu$ 
and the guiding center momenta  $Q_\mu$ ($\mu=x,y=1,2$), 
\begin{align}\nonumber
    \Pi_{\mu}& = -i\partial_\mu +e A_\mu, \\
    Q_{\mu}& = -i\partial_\mu + eA_\mu +eB\epsilon_{\mu \nu}x_\nu,
\end{align}
where $\epsilon_{\mu \nu}$ is the antisymmetric Levi-Civita symbol with $\epsilon_{12}=1$.
These operators satisfy the commutation relations 
\begin{equation}
     \left[ \Pi_\mu, \Pi_\nu \right] =- 
 \left[ Q_\mu, Q_\nu \right]= -ieB\epsilon_{\mu \nu}  , \quad \left[ \Pi_\mu, Q_\nu \right] = 0.
\end{equation}
We next define two independent sets of ladder operators,
\begin{align}\nonumber
    a^\dagger &= \frac{\Pi_x +i \Pi_y}{\sqrt{2eB}}, \quad a = \frac{\Pi_x -i \Pi_y}{\sqrt{2eB}}, \\ \label{bdef}
    b^\dagger &= \frac{(\mathbf{a}_1-i\mathbf{a}_2)\cdot \mathbf{Q}}{\sqrt{2\phi}}, \quad 
    b = \frac{(\mathbf{a}_1+i\mathbf{a}_2)\cdot \mathbf{Q}}{\sqrt{2\phi}},
\end{align}
which satisfy the canonical bosonic algebra,
$\left[ a,a^\dagger \right]  = \left[ b,b^\dagger \right] = 1$,
with all other commutators vanishing. In Eq.~\eqref{bdef}, we have introduced the magnetic
flux $\phi$ through the superlattice  unit cell (in units of $\Phi_0/2\pi$),
\begin{equation}
    \phi=eB \mathcal{A}=eBdd_y.
\end{equation}
In terms of ladder operators, the kinetic part of the Hamiltonian takes the form
\begin{equation}
   \mathcal{H}_0 =  \varepsilon_c  \begin{pmatrix} 0 & a \\ a^\dagger & 0\end{pmatrix} ,
   \label{H0}
\end{equation}
whose eigenstates are given by
\begin{equation}
    |\psi_{n, k} \rangle = \mathcal{N}_n\begin{pmatrix}
        s_{n}||n|-1,k \rangle \\
        ||n|,k \rangle
    \end{pmatrix}.  
    \label{Eqn:Landau-states1}
\end{equation}
For $\tilde n,k \in \mathbb{N}_0$, the states $|\tilde n,k\rangle$ appearing in Eq.~\eqref{Eqn:Landau-states1} are constructed as
\begin{align}
    |\tilde n,k \rangle = \frac{{a^\dagger}^{\tilde n}}{\sqrt{\tilde n!}} \frac{{b^\dagger}^{k}}{\sqrt{k!}}|0,0\rangle,
    \label{Eqn:defining-ket-states}
\end{align}
with the vacuum state defined by $a|0,0\rangle =  b|0,0\rangle = 0$.
The corresponding eigenvalues are $E_{n}  = s_{n} \sqrt{2|n|eB}$ with $n \in \mathbb{Z}$, see Eq.~\eqref{DLLfiniteM} for $M=0$.
They are highly degenerate since they do not depend on the quantum number $k$. We now exploit this degeneracy to 
construct MBSs adapted to the superlattice periodicity.

\subsection{Magnetic Bloch states}
\label{sec3b}

It is well known that in the presence of the vector potential, 
the Hamiltonian~\eqref{ham} does not commute with ordinary 
translations operators \cite{GirvinYang}.
Instead, one needs to introduce \emph{magnetic translation operators} \cite{Zak1963},
\begin{equation}
    T_{\mathbf{a}_i} =  \exp(i\mathbf{a}_i \cdot \mathbf{Q} ) .
    \label{MTOs}
\end{equation}
One easily checks that these operators commute with the kinetic Hamiltonian $\mathcal{H}_0$ in Eq.~\eqref{H0}
and with any function of position having the superlattice periodicity, 
and thus with the full Hamiltonian $\mathcal{H}$. 
Moreover,  using the Baker-Campbell-Hausdorff identity, one finds
\begin{equation}
 T_{\mathbf{a}_1}T_{\mathbf{a}_2}   = e^{-i\phi} T_{\mathbf{a}_2} T_{\mathbf{a}_1}.
\end{equation}
As a consequence, magnetic translation operators commute with one another only if $\phi$ 
is an integer multiple of $2\pi$. In a 2D superlattice, this condition is met only for discrete values of 
the magnetic field $B$, where the magnetic translation operators can be simultaneously diagonalized. 
In contrast, for our 1D mass superlattice,  it is always possible, for an arbitrary magnetic field $B$, 
to choose the transverse lattice spacing as 
\begin{equation} \label{dy}
    d_y=\frac{2\pi}{eBd},
\end{equation}
which ensures $\phi = 2\pi$.
In this case, all magnetic translation operators commute, and one can explicitly construct a basis 
of states that diagonalizes both $\mathcal{H}_0$ and all $T_{\mathbf{a}_i}$  simultaneously. 
These are the MBSs associated with the superlattice periodicity in the absence of the mass modulation. 
With a normalization factor ${\cal N}$ and the 2D quasi-momentum $\mathbf{k}$, we find
\begin{equation}
    |\mathbf{k},n \rangle = \frac{1}{\sqrt{\mathcal{N}(\mathbf{k})}} \sum_{\mathbf{R}} 
    e^{-i \mathbf{k \cdot R}} T_{\mathbf{a}_1}^{\mathbf{R} \cdot \mathbf{b}_1} 
    T_{\mathbf{a}_2}^{\mathbf{R} \cdot \mathbf{b}_2} |\psi_{n,0} \rangle ,
    \label{MBS}
\end{equation}
with the superlattice vectors $\mathbf{R}$ in Eq.~\eqref{SuperlatticeVectors} and
the Landau states $|\psi_{n,0} \rangle$ in Eq.~\eqref{Eqn:Landau-states1}.
Notice that the quantities $\mathbf{R} \cdot \mathbf{b}_i=R_i$ are integers. 
The states $|\mathbf{k},n \rangle$ in Eq.~\eqref{MBS} are eigenvectors of the magnetic translation operators,
\begin{equation}
    T_{\mathbf{a}_i}  |\mathbf{k},n \rangle = e^{i\mathbf{k} \cdot \mathbf{a}_i}  |\mathbf{k},n \rangle.
\end{equation}
The quasi-momentum $\mathbf{k}=k_1 \mathbf{b}_1+k_2 \mathbf{b}_2$ is 
defined in the first superlattice Brillouin zone with $-\pi \leq k_i< \pi$, where we recall 
 that $\mathbf{b}_2=\left(0,1/d_y \right)$ depends on the magnetic field. 
The normalization constant $\mathcal{N}(\mathbf{k})$ 
is specified in App.~\ref{AppC}, see Eq.~\eqref{Eqn:norm-const-magnetic-bloch-state}. 
Finally, one can check that the MBSs in Eq.~\eqref{MBS} satisfy the orthonormality condition \cite{Bernevig2022}
\begin{equation}
    \langle \mathbf{k}',n' |\mathbf{k},n \rangle = (2\pi)^2 \delta_{n'n} \delta (\mathbf{k}' - \mathbf{k}).
    \label{orthonormcond}
\end{equation}

\subsection{Projection on MBSs}
\label{sec3c}

We now project the full Hamiltonian $\mathcal{H}$ onto the basis of MBSs given in Eq.~\eqref{MBS}. 
Since $\mathcal{H}$ commutes with the magnetic translation operators, its matrix elements in this basis are diagonal in the quasi-momentum. 
Specifically, 
\begin{align}
    \langle \mathbf{k}',n'  | \mathcal{H} |\mathbf{k},n \rangle & = \langle \mathbf{k}',n'  |T_{\mathbf{a}_i}^\dagger 
   \mathcal{H } T_{\mathbf{a}_i}|\mathbf{k},n \rangle,  \nonumber  \\
    & = e^{i(\mathbf{k}-\mathbf{k'}) \cdot \mathbf{a}_i}  \langle \mathbf{k'},n'  | \mathcal{H}|\mathbf{k},n \rangle,
\end{align}
which implies $\langle \mathbf{k'}, n' | \mathcal{H} |\mathbf{k},n \rangle = 0$
unless $k_i -k_i' = 0 \; \mathrm{mod}\; 2\pi$. 
Hence the Hamiltonian is block-diagonal in quasi-momentum, 
and we may work at fixed $\mathbf{k}=k_1\mathbf{b}_1+k_2\mathbf{b}_2$, with $-\pi<k_i<\pi$ 
in the first Brillouin zone of the 2D superlattice. 

Eigenstates of the full Hamiltonian $\mathcal{H}$ can now be expanded in the MBS basis as
\begin{equation}
     | \Psi_{l} (\mathbf{k}) \rangle = \sum_{n\in \mathbb{Z}} c_{n}^{(l)} (\mathbf{k})| \mathbf{k},n \rangle,
     \label{Eqn:Bloch-band-eigenstate}
\end{equation}
where the sum runs only over the Landau level index $n$. 
The label $l\in \mathbb{Z}$ denotes the band index, distinguishing the exact eigenstates (Bloch bands) 
at a given quasi-momentum $\mathbf{k}$.
The diagonalization problem reduces to solving the linear system
\begin{equation}
\sum_{n\in \mathbb{Z}} \mathcal{H}_{n'n} (\mathbf{k}) \, c_{n}^{(l)}(\mathbf{k}) = \varepsilon_{l}(\mathbf{k}) \, c_{n'}^{(l)}(\mathbf{k}),
\label{linearsystem}
\end{equation}
where the matrix elements $\mathcal{H}_{n'n} (\mathbf{k})$ are defined by
\begin{align}
 \langle \mathbf{k},n'  | \mathcal{H}|\mathbf{k},n \rangle 
   = (2\pi)^2 \delta (\mathbf{0}) \mathcal{H}_{n'n} (\mathbf{k}) ,
\end{align}
  and $\varepsilon_{l}(\mathbf{k})$  denotes the exact MBBs.
The matrix elements of the kinetic term are straightforward to evaluate,
\begin{equation}
     \langle \mathbf{k},n'  | \mathcal{H}_0|\mathbf{k},n \rangle =(2\pi)^2 \delta (\mathbf{0})  E_n \delta_{n'n},
     \label{En-defined}
\end{equation}
where $E_n=s_n\sqrt{2|n|eB}$ are the $M=0$ Landau level energies.
In contrast, the periodic mass term induces inter-level scattering between different Landau levels, 
resulting in off-diagonal contributions in the index $n$. 
The calculation of these matrix elements closely follows Ref.~\cite{Bernevig2022} 
and is summarized in App.~\ref{AppC}.  Here, we only outline the key steps.

We begin by expressing the spatially modulated mass profile $m(\mathbf{r})$, 
viewed as a function of the 2D coordinate $\mathbf{r}$, as a Fourier series,
\begin{equation}
    m(\mathbf{r}) = \sum_{\mathbf{G}} A_\mathbf{G} 
    e^{-2\pi i \mathbf{G} \cdot \mathbf{r}} \;,
\end{equation}
where the sum runs over reciprocal lattice vectors 
$\mathbf{G}$, see Eq.~\eqref{reciprocallatticevectors},
and the coefficients $A_\mathbf{G}$ encode the modulation profile.
Next, we rewrite the position operator $\mathbf{r}$ in terms of magnetic ladder operators $a$ and $b$, see Eq.~\eqref{bdef}. 
The factor $e^{-2\pi i \mathbf{G}\cdot \mathbf{r}}$ then takes the form
\begin{equation}
e^{-2\pi i \mathbf{G } \cdot \mathbf{r}} = \exp \left[ -\frac{2\pi (Gb^\dagger - \overline{G}b)}{\sqrt{2\phi}}\right] 
\exp \left[ \frac{i(\overline{\gamma} a^\dagger  + \gamma a)}{\sqrt{2\phi}} \right],  
\label{Eqn:exp-G.r-simplification}
\end{equation}
where $G=G_1+iG_2$ and $\overline G$ is the complex conjugate quantity. 
With the summation convention, we also define for a wavevector $\mathbf{q}$ the quantity
\begin{equation}
    \gamma_\mathbf{q} = - \epsilon_{ij} q_i z_j ,   \label{gamma}
\end{equation}
where $q_i = \mathbf{q} \cdot \mathbf{a}_i$ and $z_i = \mathcal{A}^{-1/2}(\hat{x}+i\hat{y})\cdot\mathbf{a}_i$.   
The parameter $\gamma$ in Eq.~\eqref{Eqn:exp-G.r-simplification} is then given by $\gamma=\gamma_{2\pi \mathbf{G}}$.
The factorization in Eq.~\eqref{Eqn:exp-G.r-simplification} allows for analytical progress. Indeed,
as shown in App.~\ref{AppC}, we obtain
\begin{align}
& \langle \mathbf{k},n'  |\sigma_z e^{-2\pi i \mathbf{G} \cdot \mathbf{r}}|\mathbf{k},n \rangle = \nonumber \\
    & =(2\pi)^2 \delta(\mathbf{0}) \, e^{   i\pi G_1 G_2 -i(k_1G_2 - k_2G_1)} \, F^{(1)}_{n'n}(2\pi\mathbf{G}),
    \label{MatrixElementFC}
\end{align}
where the form factor is given by 
\begin{equation}
    F^{(1)}_{n'n}(2\pi\mathbf{G}) = 
    \mathcal{N}_{n'}\mathcal{N}_n \left[\mathrm{sgn}(n'n) 
    \mathcal{H}_{|n'|-1,|n|-1} ^{2\pi\mathbf{G}}- \mathcal{H}_{|n'|,|n|} ^{2\pi\mathbf{G}} \right],
\label{FCmatrixelement}
\end{equation}
with the matrix elements \cite{Bernevig2022}
\begin{eqnarray}
 &&   \mathcal{H}_{n'n}^{\mathbf{q}}  
     = \langle n' ,0  | \mathrm{exp} \left[  \frac{i(\overline{\gamma}_\mathbf{q} a^\dagger + 
     \gamma_\mathbf{q} a)} {\sqrt{2\phi}}\right] |n,0\rangle   
    \label{Eqn:H^q_{n'n}-definition}\\ \nonumber
    && = \left\{ \begin{array}{cc}
         e^{-\frac{\overline{\gamma}_\mathbf{q} {\gamma}_\mathbf{q}}{4\phi}}
         \sqrt{\frac{n'!}{n!}}\bigg(\frac{i\gamma_\mathbf{q}}{\sqrt{2\phi}}\bigg)^{n-n'} L_{n'} ^{(n-n')}
        \bigg(\frac{\overline{\gamma}_\mathbf{q} {\gamma}_\mathbf{q}}{2\phi}\bigg), & n\geq n', \\
        e^{-\frac{\overline{\gamma}_\mathbf{q} {\gamma}_\mathbf{q}}{4\phi}}\sqrt{\frac{n!}{n'!}}
        \bigg(\frac{i\overline\gamma_\mathbf{q}}{\sqrt{2\phi}}\bigg)^{n'-n} L_n ^{(n'-n)}
        \bigg(\frac{\overline{\gamma}_\mathbf{q} \gamma_\mathbf{q}}{2\phi}\bigg),& n<n'.
\end{array}
\right. 
\end{eqnarray}
Here, $L_n^{(n')}(x)$ are associated Laguerre polynomials \cite{NIST:DLMF}. 
Collecting the kinetic energy and mass terms, we thus arrive at the Bloch Hamiltonian matrix $\bm{\mathcal{H}}$, 
whose elements are given by
\begin{align} 
    \mathcal{H}_{n'n} (\mathbf{k}) & =  E_n \delta_{n'n} +   \label{Eqn:Hamiltonian-projection-in-magnetic-Bloch-basis}\\
   & +  \sum_{\mathbf{G}} A_{\mathbf{G}}  e^{i\pi G_1 G_2 -i(k_1G_2 - k_2G_1)} \, F^{(1)}_{n'n}(2\pi\mathbf{G}).
  \nonumber 
\end{align}
After truncation in Landau level space and Fourier space,  
numerical diagonalization of $\bm{\mathcal{H}}$
yields the exact MBBs $\varepsilon_{l}(\mathbf{k})$ as well as 
the expansion coefficients $c_n^{(l)}(\mathbf{k})$ in Eq.~\eqref{Eqn:Bloch-band-eigenstate}.
We note that for a 1D superlattice potential (independent of the $y$-coordinate), 
the Fourier components satisfy $A_\mathbf{G}\propto \delta_{G_2,0}$. Then the matrix 
elements~\eqref{Eqn:Hamiltonian-projection-in-magnetic-Bloch-basis} 
depend on $k_2$ only and the MBBs are completely flat along the periodic modulation direction ($k_x$). 
\begin{figure}[t]
\includegraphics[width=.9\linewidth]{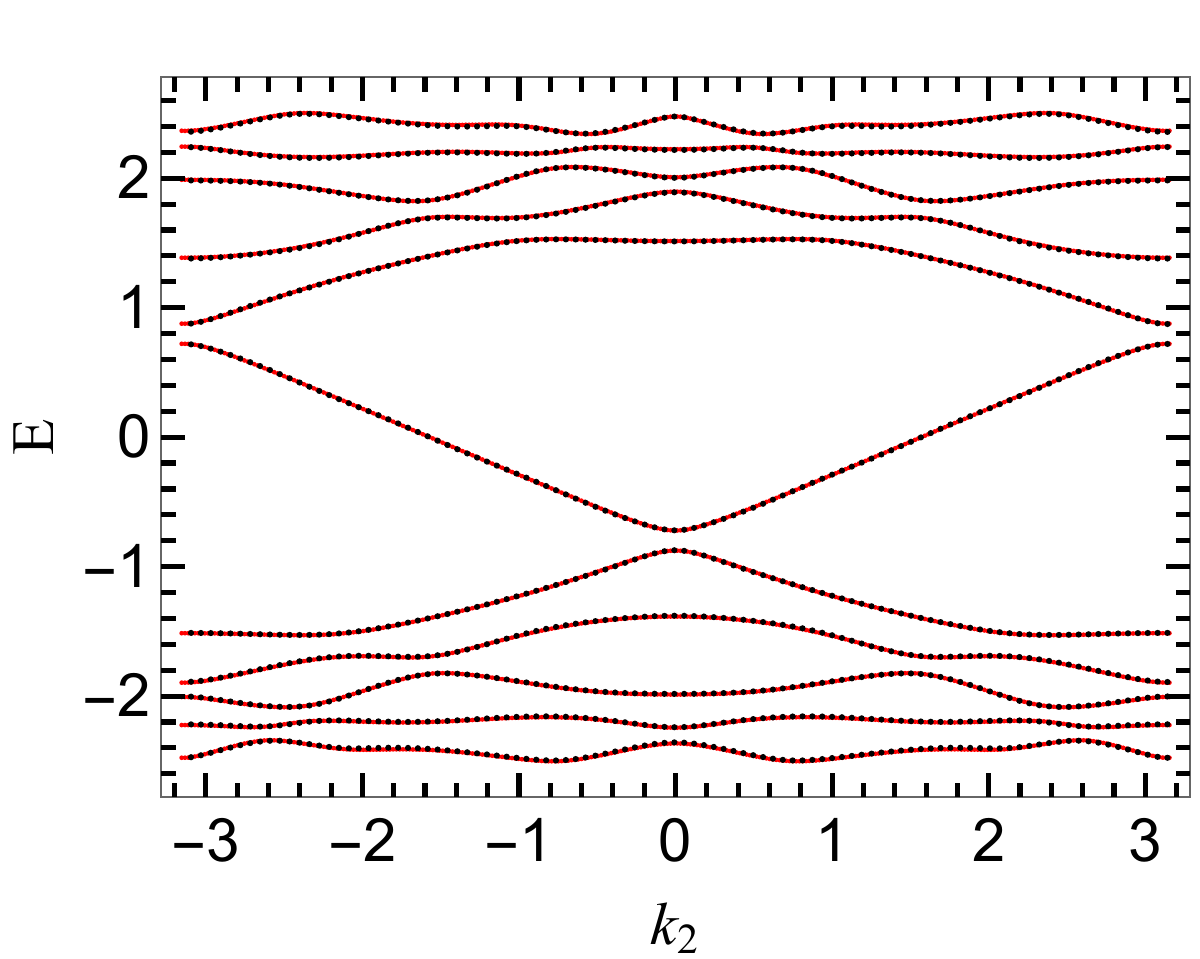}
\caption{MBBs vs $k_2 = k_y d_y\in [-\pi,\pi)$ for the periodic mass profile~\eqref{periodicprofile} 
with  $M =  \bar\varepsilon_c$ and $d = 5\bar \ell_B$ at $B=1$\,T.  
Energy is expressed in units of $\bar\varepsilon_c$.   
Black dots were obtained by numerical diagonalization of the truncated Bloch Hamiltonian 
matrix~\eqref{Eqn:Hamiltonian-projection-in-magnetic-Bloch-basis}, 
keeping the $21$ Landau states closest to zero energy and the first $200$ terms 
in the Fourier series of $m(x)$. Red curves show the energy levels of the $N=7$ array, see Sec.~\ref{sec2c}, 
for the same values of $B,M,$ and $d$, matching those used 
in Fig.~\ref{fig4}(f).
}
\label{fig6}
\end{figure}

Let us now apply this approach to the mass profile \eqref{periodicprofile}, 
which has the Fourier components
\begin{equation}
A_{\mathbf{G}}  = M (1-\delta_{G_1,0}) \delta_{G_2, 0}   \frac{\sin(G_1\pi/2)}{G_1\pi/2} .
    \label{Eqn:Fourier-coefficients-in-2D}
\end{equation}
We first verify the consistency between the spectrum obtained from exact diagonalization 
of the truncated Hamiltonian and that of a finite array of $N$ kinks and antikinks with sufficiently large $N$, see Sec.~\ref{sec2c}. 
As illustrated in Fig.~\ref{fig6}, the energy spectra 
obtained for a finite-length array with $N=7$ units (red curves) match remarkably well
to those obtained using the truncated MBS projection approach (black dots), 
thereby confirming the validity of the approach (at least for the parameters in Fig.~\ref{fig5}).

\begin{figure}[t]
    \centering
  \subfigure[]{\includegraphics[width=0.49\linewidth]{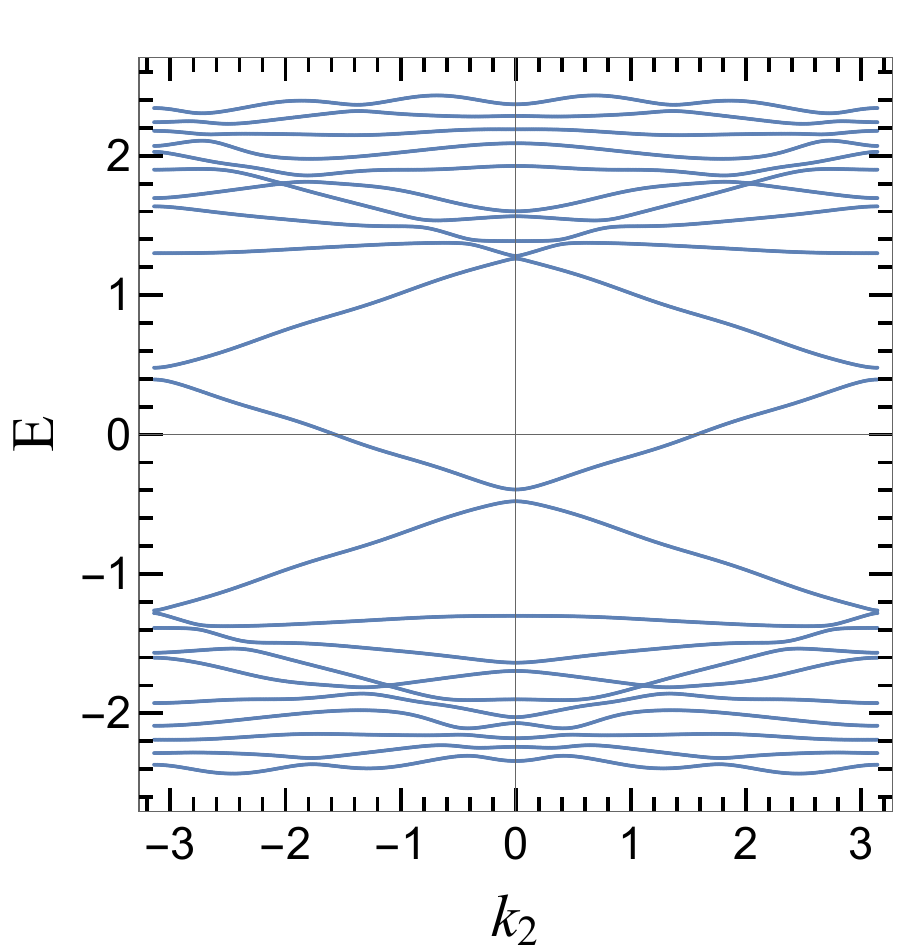}}
  \subfigure[]{\includegraphics[width=0.49\linewidth]{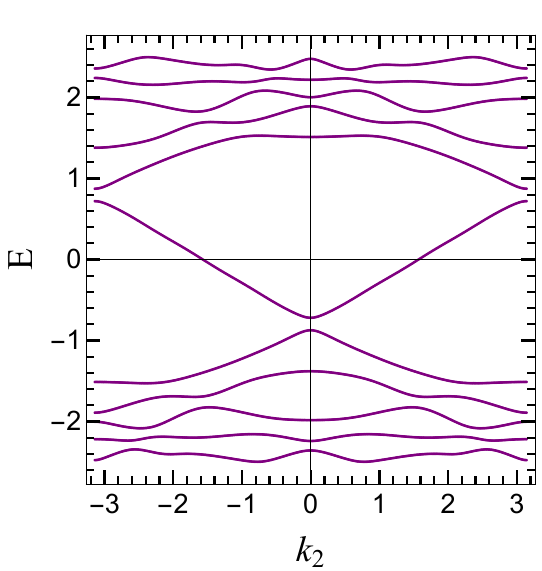}}
  \subfigure[]{\includegraphics[width=0.49\linewidth]{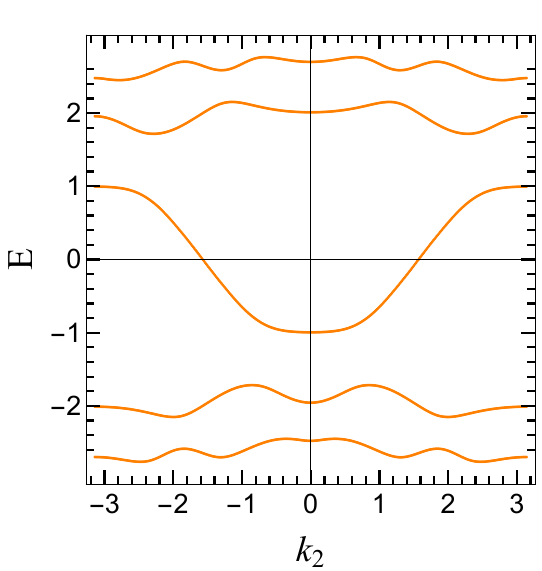}}
  \subfigure[]{\includegraphics[width=0.49\linewidth]{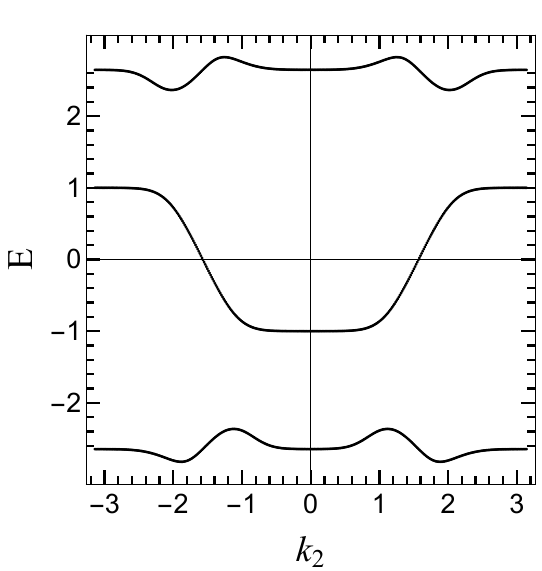}}\\
    \caption{MBBs vs $k_2 = k_y d_y\in [-\pi,\pi)$ (with energy shown in units of $\bar\varepsilon_c$) for the mass superlattice~\eqref{periodicprofile} 
    with $M = \bar\varepsilon_c$, $d = 5\bar \ell_B$, and several magnetic field values: 
    (a) $B = 0.5$~T (indigo), (b) $B=1$~T (purple), (c) $B=3$~T (orange), and (d) $B=9$~T (black curves).}
   \label{fig7}
\end{figure}

Figure~\ref{fig6} then tracks how the miniband structure evolves with magnetic field, 
highlighting the progressive narrowing of spectral gaps with decreasing field. The MBBs obey the symmetry relation 
\begin{equation}
\varepsilon_l(k_2\pm \pi) = -\varepsilon_{-l}(k_2),
\end{equation}
which follows directly from the property $m(x+d/2)=-m(x)$. Combined with the inversion symmetry 
$\varepsilon_l(-k_2) = \varepsilon_{l}(k_2)$, 
this enforces the $l=0$ level to cross zero exactly at $k_2=\pi/2$.
We note that the apparent steepening of the chiral branches at larger $B$
is not physical, but results from the widening of the Brillouin zone along the $k_y$-direction.
The plotted interval  $|k_2|< \pi$ in fact maps to the interval $|k_y|<\pi/d_y$, which broadens as $B$ increases.
This effect causes the slope of the chiral mode dispersions near zero energy in Fig.~\ref{fig6}   
to appear steeper with increasing $B$, although the magnetic field actually 
suppresses the mode velocity, as discussed in see Sec.~\ref{sec2b}.   

\section{Transport observables}
\label{sec4}

Using the results of Secs.~\ref{sec2} and~~\ref{sec3}, we now investigate the transport 
signatures of Dirac fermions subject to a mass superlattice and a uniform magnetic field.
In Sec.~\ref{sec4a}, we study the Hall conductivity, and in Sec.~\ref{sec4b}, 
we discuss Weiss-like magnetoconductivity oscillations induced by the mass modulation.

\subsection{Hall conductivity}
\label{sec4a}

Within linear response theory, the Hall conductivity follows from the Kubo formula as \cite{thouless1982, Ong-Anomalous}
\begin{equation}\label{Eqn:Hall-cond.}
    \sigma_{xy} = e^2 \hbar\sum_{\substack{l, l' \\ l \neq l'}} \int \frac{d\mathbf{k}}{(2\pi)^2} [f(\varepsilon_{l}(\mathbf{k})) - 
    f(\varepsilon_{l'}(\mathbf{k}))] \;\Omega^{xy}_{l,l'}(\mathbf{k})
\end{equation}
with the Berry curvature 
\begin{equation}
    \Omega^{xy}_{l,l'}(\mathbf{k}) =  \mathrm{Im} \frac{\langle \Psi_{l} (\mathbf{k})| v_x | \Psi_{l'} (\mathbf{k}) \rangle 
    \langle \Psi_{l'} (\mathbf{k})| v_y | \Psi_{l} (\mathbf{k}) \rangle}{[\varepsilon_{l}(\mathbf{k}) - \varepsilon_{l'}(\mathbf{k})]^2}.
\end{equation}
Here, $|\Psi_{l} (\mathbf{k}) \rangle$ and $\varepsilon_{l}(\mathbf{k})$ 
are the exact MBSs and MBBs in Eqs.~\eqref{Eqn:Bloch-band-eigenstate} and~\eqref{linearsystem}, 
respectively, $f(\varepsilon)$ is the Fermi distribution function, and $v_{x,y}=v_\mathrm{F}\sigma_{x,y}$ 
are velocity operators. For a 1D mass superlattice, the Bloch Hamiltonian 
matrix~\eqref{Eqn:Hamiltonian-projection-in-magnetic-Bloch-basis}
and the resulting MBBs are independent of the wave vector component $k_x$ along the superlattice direction, and thus the $\mathbf{k}$-dependence appears only via $k_y$. 
Consequently, Eq.~\eqref{Eqn:Hall-cond.} reduces to an integral over $k_y$, 
which we evaluate numerically using the MBS projection approach in Sec.~\ref{sec3c}.

Specifically, we expand $|\Psi_l(\mathbf{k})\rangle$ using the complex-valued MBS expansion coefficients 
$c_n^{(l)}(\mathbf{k})$ in Eq.~\eqref{Eqn:Bloch-band-eigenstate} and compute the expectation values of the velocity operators.
After some algebra, including a change of integration variables from $(k_x,k_y) \to (k_1,k_2)$ 
with $k_i = \mathbf{k} \cdot \mathbf{a}_i$ and an integration over $k_1$,   we arrive at
\begin{equation}
\sigma_{xy} =\frac{e^2B}{4\pi hB_0} \int_{-\pi}^{\pi} \!\! dk_2 
\sum_{\substack{ l \neq l'}}
\left[ f(\varepsilon_{l}(k_2)) - f(\varepsilon_{l'}(k_2))\right ] 
\tilde{\Omega}^{xy}_{l,l'}(k_2), 
      \label{Eqn:Main-expression-simplified-Hall-conductivity}
\end{equation}
where $B_0=1$\,Tesla. The dimensionless Berry curvature is

\begin{equation}
    \label{berry2}
   \tilde{\Omega}^{xy}_{l,l'} (k_2)  =
   -\frac{\mathrm{Re} \mathlarger{\sum \limits_{\substack{n_1,n_2,\\ n_3, n_4}}} c_{n_1}^{(l)\,*}c_{n_2}^{(l')} c_{n_3}^{(l')\,*}c_{n_4}^{(l)} 
    F_+ (n_1,n_2) F_- (n_3,n_4) }{\left[ \varepsilon_{l}(k_2) - \varepsilon_{l'}(k_2) \right]^2/\bar \varepsilon_c^2},
\end{equation}
where a $k_2$-dependence also appears via the coefficients $c_n^{(l)}$.
%We here omit an overall factor $(2\pi)^2\delta(\mathbf{0})$ in the matrix elements, which is absorbed by normalization constants 
%when converting summations over discrete states into integrals in Eq.~\eqref{Eqn:Hall-cond.}.
With $\mathcal{N}_n$ in Eq.~\eqref{wf}, the functions $F_{\pm}(n_a,n_b)$ are given by
\begin{equation}
    F_{\pm} =  \mathcal{N}_{n_a}\mathcal{N}_{n_b} \left[\mathrm{sgn}(n_a) \delta_{|n_a|-1,|n_b|}\pm\mathrm{sgn}(n_b) \delta_{|n_a|,|n_b|-1} \right]. 
\end{equation}

\begin{figure}[t]
    \centering
    \includegraphics[width=8cm]{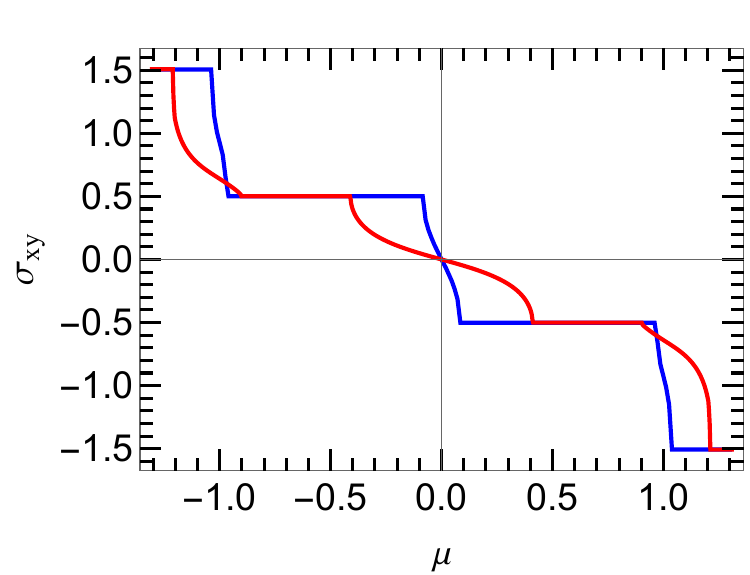}
    \caption{Zero-temperature Hall conductivity (in units of $e^2/h$)  vs chemical potential $\mu$ (in units of $\varepsilon_c$) 
    for Dirac fermions in the mass superlattice \eqref{periodicprofile} at $B=1$\, Tesla.  
    We use the lattice period $d=5 \bar \ell_B$, 
    and the superlattice amplitudes (a) $M = 0.1 \bar \varepsilon_c$ (in blue), and (b) $M = 0.5 \bar \varepsilon_c$ (in red).} 
    \label{fig8}
\end{figure}
Assuming the specific step-like form in Eq.~\eqref{periodicprofile} for the 1D mass superlattice, 
with amplitude $M$ and period $d$, we evaluate the above expressions numerically.  
We show the resulting $T=0$ Hall conductivity for $d/\ell_B=5$ 
as function of the chemical potential $\mu$ in Fig.~\ref{fig8}.
For very strong fields, $M\ll \varepsilon_c$, MBBs basically reduce to flat Landau levels with 
$\varepsilon_l (k_2)  \approx s_lv_\text{F} \sqrt{2\hbar|l|B}$ and $c_{n}^{(l)} \approx \delta_{n,l}$.
For chemical potentials with $0<|\mu|<\varepsilon_c$, 
we then recover from Eqs.~\eqref{Eqn:Main-expression-simplified-Hall-conductivity}
and \eqref{berry2} the well-known half-integer massless graphene sequence (without spin-valley degeneracy) \cite{Gusynin2005}, 
$\sigma_{xy} = -{\rm sgn}(\mu) \frac{e^2}{2 h}$, consistent with the small-$M$ result in Fig.~\ref{fig8}.

For larger values of $M/\varepsilon_c$, however, the mass superlattice
hybridizes the Landau levels and reshapes the spectral minigaps. 
While quantized Hall plateaus still persist, see Fig.~\ref{fig8}, they become narrower in width (with respect to the chemical potential) 
since the Landau-level mixing by the superlattice reduces the spectral gaps between different MBBs and enhances the band dispersion, 
see Fig.~\ref{fig6}.  Such effects are clear manifestations of the mass superlattice in the Hall conductivity, 
destroying the half-integer quantization of $\sigma_{xy}$ for certain chemical potential regimes.

\subsection{Weiss oscillations}
\label{sec4b}

Next, we turn to Weiss oscillations, which are commensurability oscillations of the 2D magnetoconductivity. In a semiclassical picture, 
such oscillations arise from the interplay between the cyclotron motion (at the Fermi energy) and a weak spatially periodic potential modulation \cite{Weiss1989,Winkler1989,Gerhardts1989,Beenakker1989,Pfannkuche1992,Peeters1992,Beenakker1989}. In graphene and other Dirac systems, 
Weiss oscillations have been predicted and observed for electrostatic superlattices \cite{Matulis2007,Eroms2018,Dean2021,Huber2022,Paul2022,Koop2024}, 
with an amplitude enhancement compared to the conventional case of a 2D electron gas. The characteristic periodicity in $1/B$ 
is set by the superlattice period $d$ and the Fermi wave number $k_\text{F}$ via the commensurability condition  \cite{Weiss1989}:
\begin{equation}
    2R_c = \left( \lambda - \frac{1}{4} \right) d, \quad \lambda = 1,2,3,\ldots,
\end{equation}
where $R_c = \frac{\hbar k_F}{eB}$ is the cyclotron radius.

Below, we show that a weak-amplitude mass superlattice $m(x)$ will also cause Weiss-like magnetoconductivity oscillations 
in $\sigma_{yy}$ for 2D Dirac fermions. Compared to the corresponding electrostatic case \cite{Matulis2007,tahir2007}, 
we predict a strongly reduced oscillation amplitude with a robust phase shift $\pi/2$.
We follow the standard semiclassical Drude–Boltzmann approach \cite{Matulis2007}. 
To that end, we first obtain the energy shifts of the Landau levels using first-order perturbation theory in $m(x)$. 
In the second step, we insert the resulting group velocities into the semiclassical expression for $\sigma_{yy}$. 

Assuming that the mass profile $m(x)$ is even in $x$, we start by expanding it as a cosine series,
\begin{equation}
       m(x)  = \sum_{j=1}^\infty m_j \cos \left(\frac{2\pi j x}{d} \right).
       \label{cosinemass}
\end{equation}
Applying first-order perturbation theory in $m(x)$, the correction to  the Landau level energies $E_n$ (given by 
Eq.~\eqref{DLLfiniteM} for $M=0$) 
takes the form (see App.~\ref{AppD})
\begin{align}\nonumber
   \Delta E_{n,k_y}
   &=\langle \Psi_{n,k_y}| m(x)\sigma_z |\Psi_{n,k_y}\rangle \\
         &= \mathcal{N}_{n}^2\sum_j m_j \cos \left(jk_yd_y  \right) e^{-\frac{1}{2}j^2u} \times \label{Eqn:Delta-E-weiss-osc.} \\ \nonumber 
         & \times \left[ (1-\delta_{n,0})  L_{|n|-1} (j^2u)-L_{|n|}(j^2u)  \right],        
\end{align}
with the Landau spinors $|\Psi_{n,k_y}\rangle$ defined in  Eq.~\eqref{wf}
and $L_n (x)$ the Laguerre polynomials \cite{NIST:DLMF}. 
We again use $d_y=2\pi \ell_B^2/d$, see Eq.~\eqref{dy}, and employ the dimensionless variable 
$u = \frac{d_y ^2}{2\ell_B^2}=2\pi^2 \frac{B_W}{B}$, where $B_W=\frac{\hbar}{ed^2}$ 
is the magnetic field at which the magnetic length $\ell_B$ equals the superlattice period $d$.
The drift velocity along the $y$-direction is then given by
\begin{eqnarray}\nonumber
&&    v_y^{n,k_y}= \frac{1}{\hbar} \frac{\partial}{\partial k_y} \Delta E_{n,k_y}= 
- \frac{\mathcal{N}_{n}^2}{\hbar}\sum_j j d_y m_j
    \sin \left(jk_yd_y  \right) \\  \label{velocity1}
         && \quad \times  e^{-\frac{1}{2}j^2u} \left[ (1-\delta_{n,0})  L_{|n|-1}(j^2u)-L_{|n|}(j^2u)  \right].
\end{eqnarray}
The diffusive longitudinal conductivity is given by~\cite{Matulis2007} 
\begin{align}\label{diffcond}
    \sigma_{yy} = \frac{e^2 \beta}{L_x L_y} \sum_{{\zeta}}
f(E_{\zeta}) \left[ 1 - f(E_{{\zeta}}) \right] \tau(E_{{\zeta}})
\left( v_{ y}^{{\zeta}} \right)^2,
\end{align}
where $\zeta$ runs over all quantum numbers and $\beta=1/k_B T$.

Let us now focus on a simple cosine mass profile, where $m_j = m_1 \delta_{j,1}$. 
Assuming a smooth energy dependence of the electron relaxation time 
$\tau(E)\approx \tau(E_{\rm F})=\tau$ due to  electron-impurity scattering processes 
(not explicitly included in our model), see Ref.~\cite{Matulis2007},
and carrying out the $k_y$-integration, we arrive at  
\begin{equation}\label{sigmayy}
    \sigma_{yy} = \sigma_0 \Phi, \quad 
    \sigma_0 =  \frac{e^2}{h}\frac{m_1^2 \beta \tau}{\hbar},
\end{equation}
with the dimensionless conductivity 
\begin{equation}\label{Eqn:dimensnless-cond-weiss}
   \Phi = ue^{-u} \sum_{n = -\infty}^{+\infty} 
   \frac{\mathcal{N}_n^4 \left[ (1-\delta_{n,0}) L_{|n|-1}(u)-L_{|n|}(u)  \right]^2}{4\cosh^2[\beta(E_n-\mu)/2]}.
\end{equation}
For the generalization  to mass profiles containing many Fourier components, see App.~\ref{AppD}.

\begin{figure}[t]
    \centering
    \includegraphics[width=7cm]{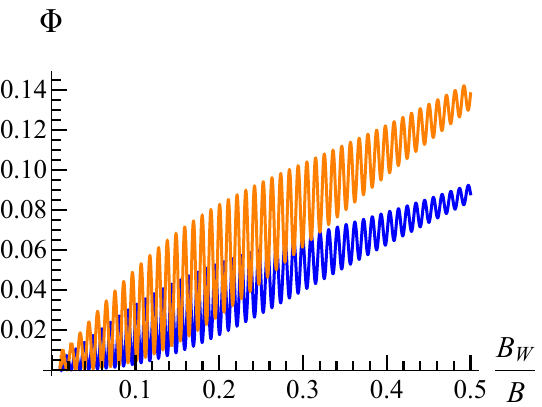}
    \caption{Weiss oscillations of the magnetoconductivity for 2D Dirac fermions. We show the dimensionless magnetoconductivity, $\Phi=\sigma_{yy}/\sigma_0$, see 
    Eqs.~\eqref{sigmayy} and \eqref{Eqn:dimensnless-cond-weiss}, as function of the inverse magnetic field for a weak mass superlattice $m(x)=m_1\cos(2\pi x/d)$. 
    The magnetic field $B_W = \frac{\hbar}{e d^2}$ is defined after Eq.~\eqref{Eqn:Delta-E-weiss-osc.}. 
    The result is shown as blue curve and scaled up by a factor $150$ for visibility. The orange curve illustrates the corresponding results 
    for an electrostatic superlattice with the same amplitude and period~\cite{Matulis2007}. 
    We employ the parameters $T = 6\,$K, $n_e = 1.5 \times 10^{11}\mathrm{cm}^{-2}$, and $d = 350\,$nm.  
    The Fermi energy for the quoted electron density $n_e$ is $90$\,meV.} 
    \label{fig9}
\end{figure}

Equation~\eqref{Eqn:dimensnless-cond-weiss} is the mass-modulation analogue of the electrostatic case considered in Ref.~\cite{Matulis2007}. 
The only structural change is the relative minus sign between the Laguerre polynomials in Eq.~\eqref{Eqn:dimensnless-cond-weiss}, 
which ultimately originates from the opposite coupling of the two spinor components to $m(x)\sigma_z$. 
This minus sign implies a strongly reduced amplitude and a phase difference of $\pi/2$ 
for Weiss oscillations in a  mass superlattice as compared to the electrostatic case \cite{Matulis2007}.  
Specifically, in the semiclassical limit of high Landau indices $|n|\gg 1$ and moderate values of $u\propto 1/B$, 
the asymptotic behavior of the Laguerre polynomials \cite{NIST:DLMF} implies
\begin{equation}
 e^{-\frac{u}{2}}   \left( L_{|n|-1}(u) - L_{|n|}(u) \right) 
 \sim \frac{u}{\sqrt{\pi}(nu)^{\frac{3}{4}}} \sin\left(2\sqrt{nu} -\frac{\pi}{4} \right) ,
\end{equation}
whereas for the electrostatic case \cite{Matulis2007} featuring the sum of both terms, one finds
\begin{equation}
 e^{-\frac{u}{2}}   \left( L_{|n|-1}(u) + L_{|n|}(u) \right) \sim \frac{2}{\sqrt{\pi}(nu)^{\frac{1}{4}}} \cos\left(2\sqrt{nu} -\frac{\pi}{4} \right) .
\end{equation}
The $\sin$ versus $\cos$ dependence makes the $\pi/2$ phase shift explicit, while the extra factor $\sqrt{u/4n}$ suppresses the mass-case amplitude compared to the electrostatic result.
After squaring and summing over thermally broadened Landau levels
in Eq.~\eqref{Eqn:dimensnless-cond-weiss}, we obtain oscillations in $\Phi(1/B)$ 
which are shifted by $\pi/2$ and strongly reduced in amplitude for the mass modulation case, as
illustrated in Fig.~\ref{fig9}.

Finally, for a multi-harmonic mass superlattice, e.g., the one in Eq.~\eqref{periodicprofile}, 
each Fourier component $m_j$ generates an oscillatory contribution of the same form as Eq.~\eqref{Eqn:dimensnless-cond-weiss}  
with $u\rightarrow j^2u$. The superposition of these components produces beating and/or aperiodic patterns in $\Phi(1/B)$, 
in particular when several $m_j$ are of comparable size, see App.~\ref{AppD} for details.

\section{Conclusions}
\label{sec5}

In this paper, we have combined exact solutions based on the transfer matrix technique, a low-energy formulation, and a gauge-invariant MBS projection approach
in order to study 2D Dirac fermions in a 1D periodic mass potential $m(x)$ under the influence of a perpendicular magnetic field $B$. 
For isolated domain walls in the mass profile, we find that the magnetic field renormalizes the velocity 
of chiral Jackiw-Rebbi modes confined to the mass kink and unidirectionally propagating along the $y$-direction. 
In arrays of $N$ kink-antikink pairs, the coupling between these interface states produces dispersive minibands 
whose structure depends sensitively on the kink-kink distance relative to the magnetic length.
For the infinite periodic mass superlattice case, the MBS projection method is more efficient 
and yields the full miniband spectrum for arbitrary periodic mass profile $m(x)$.
We already find excellent agreement between the results of the MBS projection approach 
and those of finite-size array calculations for $N=7$ and otherwise identical parameters. 
From these spectra we conclude that mass superlattices modify the width of quantum Hall plateaus and, 
using a complementary perturbative treatment, that they give rise to Weiss-type magnetoconductivity oscillations. 
Compared with the standard electrostatic superlattice case, these oscillations exhibit a strongly reduced amplitude 
and a distinct $\pi/2$ phase shift.

Our theoretical results demonstrate how periodic mass modulations can reshape the interplay between magnetic quantization and 
Dirac fermion dynamics.  Such modulations offer a general framework for miniband engineering of Dirac materials, in particular, 
for graphene monolayers or the surface states in  topological insulators.  Based on the results reported above,  
one may design superlattice-based devices that exploit magnetic miniband engineering.  
For instance, one can prepare chiral 1D Jackiw-Rebbi modes with arbitrary velocity (below the Fermi velocity). 
\new{We expect many of the qualitative findings reported here to carry over to  multilayer graphene systems.  
Given the current interest in such setups, a detailed investigation of this case remains a promising direction for future research.} 

\new{We hope our work will motivate further theoretical and experimental studies along these lines.}

\begin{acknowledgments}
We thank D. Bercioux, L. Dell'Anna, and C. Mora for useful discussions. AA is grateful for the PhD studentship provided by City St George's, University of London. R.E.~acknowledges funding by the Deutsche Forschungsgemeinschaft (DFG, German Research Foundation) 
under Projektnummer 277101999 - TRR 183 (projects B02) and under Germany's Excellence Strategy - 
Cluster of Excellence Matter and Light for Quantum Computing (ML4Q) EXC 2004/1 - 390534769.
\end{acknowledgments}

\section*{Data availability}
The data underlying the figures in this paper are available in Zenodo \cite{Zenodo}.

%%%%%%%%%%%%%%%%%%%%%%%%%%%%%%%%%%%%%%%%%%%%%%%%%%%%%%%%%%%%%%%%%
% APPENDIX
%%%%%%%%%%%%%%%%%%%%%%%%%%%%%%%%%%%%%%%%%%%%%%%%%%%%%%%%%%%%%%%%%

\appendix
\setcounter{figure}{0}
\renewcommand{\thefigure}{A\arabic{figure}}

\section{Smooth mass kink}
\label{AppA}

\begin{figure}[t]
  \centering
  \subfigure[]{\includegraphics[width=0.49\linewidth]{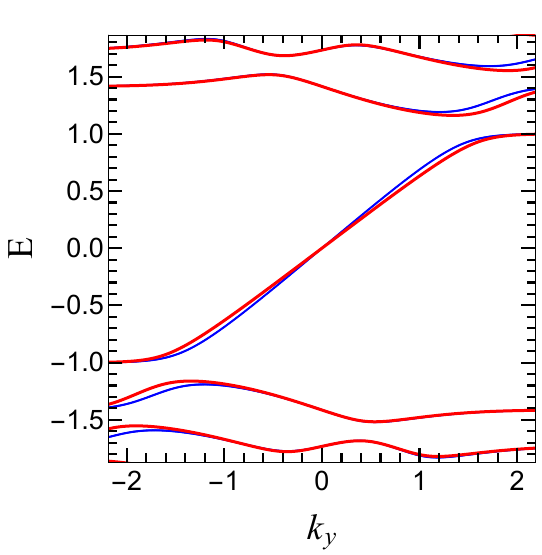}}
  \subfigure[]{\includegraphics[width=0.49\linewidth]{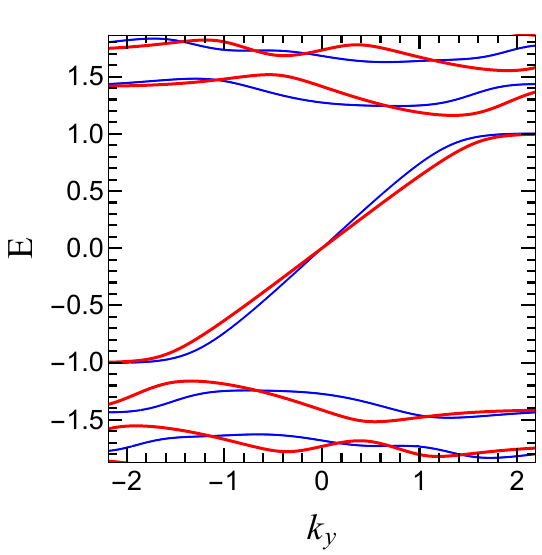}}
\caption{Energy spectra of 2D Dirac fermions in a magnetic field $B=1$~T
for a piecewise linear mass kink profile with length scale $\ell$ (see Eq.~\eqref{piecewisemass}, blue curves) and for a sharp kink profile ($\ell=0$, see Eq.~\eqref{kink}, red curves). We study two cases: (a) $\ell = 0.1\bar \ell_B$ and (b) $\ell = \bar\ell_B$.  Energies and wave vectors are given in  
units of $\bar \varepsilon_c$ and $\bar \ell_B^{-1}$, respectively, see Eq.~\eqref{scalesdef2}.}
\label{figApp1}
\end{figure}

Here we calculate the spectrum of the Hamiltonian~\eqref{ham} for an extended (smooth) mass kink profile with a characteristic length scale $\ell$. 
To keep the problem exactly solvable, we consider a piecewise linear mass profile,
\begin{equation}
    m(x) = \begin{cases}
         - M, &   x < - \ell \\
        Mx/\ell, &  |x| \leq \ell \\
        + M,    &  x > \ell
    \end{cases} .
    \label{piecewisemass}
\end{equation}
For $|x|>\ell$, the wave function reads
\begin{equation}
  \psi(x) = \begin{cases}
  W_{-M}(x) \begin{pmatrix} 
a_L \\ 0
\end{pmatrix}   &  \text{for} \; x<-\ell, \\
W_{M}(x) \begin{pmatrix} 
0 \\ b_R
\end{pmatrix}  &  \text{for} \; x>\ell, \\
  \end{cases}   
\end{equation}
with $W_M(x)$ in Eq.~\eqref{WmatrixApp}.
For $|x| \leq \ell$, the slope of the 
mass profile effectively renormalizes the magnetic field.
Using the results of App.~\ref{AppB}, with $m_0=0$ and $m'=M/\ell$ in Eq.~\eqref{linmass}, 
the wave function reads
\begin{equation}
  \psi( x) =  U^{-1} \, W_\ell( x) \begin{pmatrix} 
c_1 \\ c_2
\end{pmatrix},  
\end{equation}
where the unitary matrix $U$ and the corresponding angle $\alpha$ are defined in App.~\ref{AppB}.
Moreover, the matrix $ W_\ell(x)$ is given by 
\begin{equation}
 W_\ell(x) = 
    \begin{pmatrix}    \frac{(E-k_y\sin\alpha)}{\widetilde \varepsilon_c} D_{\widetilde p-1}(-\widetilde q) & 
    \frac{( E- k_y \sin\alpha)}{\widetilde \varepsilon_c} 
    D_{\widetilde p-1}(\widetilde q)  &  \\
     -iD_{\widetilde p}(-\widetilde q) & iD_{\widetilde p}(\widetilde q)  \end{pmatrix},
     \label{Wmatrixren}
\end{equation}
with the quantities
$$
\widetilde p= \frac{E^2-k_y^2\sin^2\alpha}{\widetilde \varepsilon_c},\quad 
\widetilde q= \frac{\sqrt{2}( x +  k_y \cos \alpha)}{\widetilde \ell_B}, 
$$
where  $\widetilde \varepsilon_c=\sqrt{2e\widetilde B}$ and $\widetilde \ell_B=1/\sqrt{e \widetilde B}$, 
see Eq.~\eqref{renB} for the definition of  $\alpha$ and $\widetilde B$.
Since the region of linear mass extends from $-\ell$ to $\ell$, 
we impose continuity of the wave function at $x=\pm \ell$,
\begin{align}
W_{-M}(-\ell)\begin{pmatrix}  a_L \\ 0 \end{pmatrix} 
& = U^{-1}W_\ell (-\ell) \begin{pmatrix} c_1 \\ c_2 \end{pmatrix}  , \\ \nonumber
 U^{-1}W_\ell(\ell)
\begin{pmatrix} c_1 \\ c_2 \end{pmatrix} 
&= W_{M}(\ell) \begin{pmatrix}  0 \\ b_R \end{pmatrix}.
\end{align}
We then find the relation
$\begin{pmatrix}  0 \\ b_R \end{pmatrix} = \bm{\Omega}  \begin{pmatrix}  a_L \\ 0 \end{pmatrix}$,
with the transfer matrix
\begin{equation}
\bm{\Omega} = W_M^{-1}(\ell) \, U^{-1} \, W_\ell(\ell) \,  W_\ell^{-1}(-\ell) \, U \, W_{-M}(-\ell).
\end{equation}
The spectrum follows from the condition ${\Omega}_{11} = 0$
and is shown in Fig.~\ref{figApp1}.
While the smooth domain wall introduces an additional length scale 
$\ell$, the resulting energy levels retain the same qualitative features as those obtained 
for the sharp-kink profile in Sec.~\ref{sec2b}.

\section{Linear mass profile}
\label{AppB}

We here provide the exact solution of the 2D Dirac equation with the linear mass profile 
\begin{equation}
\label{linmass}
  m(x)=m_0+m' x,  
\end{equation} 
in a constant magnetic field. 
The corresponding problem of a linear mass profile with a constant electric field has been solved in Ref.~\cite{Yuhong1984}, 
see also Ref.~\cite{Lukose2007}.  Those results are used in Sec.~\ref{sec2d} and in App.~\ref{AppA}.

To diagonalize $\mathcal{H}(k_y)$ in Eq.~\eqref{ham} for a linear mass profile, 
we first apply a unitary transformation,  $U = e^{i \frac{\alpha}{2} \sigma_x }$, with 
\begin{equation} 
\cos \alpha = \frac{B}{\widetilde B},\quad \sin\alpha = \frac{m'}{e\widetilde B}, \quad  e \widetilde  B = \sqrt{(eB)^2+m'^2}.
\label{renB}
\end{equation}
Using 
$ U \left( eB \sigma_y + m' \sigma_z \right) U^{-1}  =e\widetilde B \sigma_y,$
the transformed Hamiltonian $\widetilde {\mathcal H} (k_y)   = U \mathcal H (k_y) U^{-1 }$ reads 
\begin{align}
\widetilde     {\mathcal H} (k_y)  =
    \sigma_x \hat p_x  +  \left( e\widetilde B x +  \widetilde k_y \right) \sigma_y  + \widetilde m_0 \sigma_z,% + \mathcal{E}x,
  \label{hamrotated}
\end{align}
where we have defined
\begin{align}
   \widetilde k_y &=k_y \cos\alpha + m_0 \sin\alpha ,  \nonumber \\
    \widetilde m_0 &= -k_y\sin\alpha + m_0 \cos\alpha. \label{m0}
\end{align}
Equation~\eqref{hamrotated} is a standard massive Dirac Hamiltonian in a perpendicular magnetic field, where the strength 
of the field is renormalized by the finite slope $m'$ of the mass profile. The energy levels are thus given by
\begin{align}
  E_0(k_y)=-\widetilde m_0,\quad  E_{n}(k_y) = s_n \sqrt{ 2 e \widetilde B|n|+ \widetilde m^2_0 },  \quad 
    n \in \mathbb{Z}^\ast.
\end{align}
These energy levels are dispersive because  $\widetilde m_0$ depends on $k_y$, see Eq.~\eqref{m0}\\

\section{Matrix elements}
\label{AppC}

Here we summarize the key steps in the
calculation of the matrix elements~\eqref{MatrixElementFC}, see Ref.~\cite{Bernevig2022}
for an in-depth discussion. We first recall that our construction 
is based on an expansion in Landau states, as implicitly considered in Ref.~\cite{Bernevig2022} as well.
Using the magnetic translation operators $T_{\mathbf{a}_i} = \mathrm{exp}(i\mathbf{a}_i \cdot \mathbf{Q})$, 
the MBSs~\eqref{MBS} can be cast in the form
\begin{equation*}
    |\mathbf{k},n \rangle = \frac{1}{\sqrt{\mathcal{N}(\mathbf{k})}} \sum_{\mathbf{R}} 
    e^{-i \mathbf{k \cdot R}+i\frac{\phi}{2}R_1 R_2} e^{i \mathbf{R \cdot Q}} |\psi_{n,0} \rangle.
\end{equation*}
We want to evaluate the matrix elements
\begin{widetext}
\begin{equation}
      \langle \mathbf{k},n'  | \sigma_z e^{-2\pi i \mathbf{G \cdot r}} |\mathbf{k},n \rangle = 
      \frac{1}{{\mathcal{N}(\mathbf{k})}}  \sum_{\mathbf{R},\mathbf{R}'} e^{-i \mathbf{k} \cdot (\mathbf{R}-\mathbf{R}')} 
      e^{-i\frac{\phi}{2} R'_1 R'_2 + i\frac{\phi}{2} R_1 R_2} \langle \psi_{n',0} | 
      \sigma_z e^{ -2\pi i \mathbf{G} \cdot \mathbf{r} } e^{ -i\mathbf{R}' \cdot \mathbf{Q} } e^{ i\mathbf{R} \cdot \mathbf{Q} } |\psi_{n,0} \rangle,
      \label{Eqn:Ber-matrix-elt-01}
\end{equation}
\end{widetext}
where $e^{-2\pi i \mathbf{G} \cdot \mathbf{r}}$ commutes with the magnetic translation operators.
To simplify $\mathbf{G} \cdot \mathbf{r}$, we represent the position operator in terms of  the ladder operators $b$ and $a$ in Eq.~\eqref{bdef},
\begin{eqnarray*}
   \mathbf{b}_1 \cdot \mathbf{r}  &  =& -\frac{1}{\sqrt{2\phi}} \left[ i \left(b^\dagger - b) -  (z_2 a + \overline{z}_2 a^\dagger\right) \right], \\
   \mathbf{b}_2 \cdot \mathbf{r}  &=& \frac{1}{\sqrt{2\phi}} \left[ \left(b^\dagger + b) -  (z_1 a + \overline{z}_1 a^\dagger\right) \right],
\end{eqnarray*}
with $z_i = \mathcal{A}^{-1/2}(\hat{x}+i\hat{y})\cdot\mathbf{a}_i$. 
As a result, we arrive at Eq.~\eqref{Eqn:exp-G.r-simplification}.

To proceed further, we also write $\mathbf{R} \cdot \mathbf{Q}$ in terms of  $b$,
\begin{equation} 
    e^{i\mathbf{R} \cdot \mathbf{Q}}  = 
    e^{-\frac{\phi}{4}\overline{R}R} \;  e^{i\sqrt{\frac{\phi}{2}} R b^\dagger}\;  e^{i\sqrt{\frac{\phi}{2}}\;\overline{R} b} ,
    \label{Eqn:exp-R.Q-simplification}
\end{equation}
where $R = R_1+iR_2$. Note that $[ \mathbf{R} ' \cdot \mathbf{Q}, \mathbf{R} \cdot \mathbf{Q}  ] = i\phi (R_1' R_2 -R_2' R_1).$
Using the Baker-Campbell-Hausdorff formula and Eqs.~\eqref{Eqn:exp-G.r-simplification} and  \eqref{Eqn:exp-R.Q-simplification},
we obtain
\begin{eqnarray*}
    &&  \langle \psi_{n',0} |\sigma_ze^{-2\pi i \mathbf{G } \cdot \mathbf{r}} e^{i\mathbf{R} \cdot \mathbf{Q}}|\psi_{n,0} \rangle 
        = e^{-\frac{(2\pi)^2}{4\phi}\overline{G}G-\frac{\phi}{4}\overline{R}R+i\pi \overline{G}R} \times \\ && \quad \times
        \langle \psi_{n',0} |\sigma_z \;e^{\frac{i}{\sqrt{2\phi}}(\overline{\gamma}a^\dagger + {\gamma}a)}  \; |\psi_{n,0} \rangle.          
\end{eqnarray*}
Next, using the matrix elements
\begin{eqnarray*}
 &&   \langle \psi_{n',0} |\sigma_z \;e^{\frac{i}{\sqrt{2\phi}}(\overline{\gamma}a^\dagger + {\gamma}a)} |\psi_{n,0} \rangle = 
    \mathcal{N}_n \mathcal{N}_{n'} \times \\ 
&& \biggl[\mathrm{sgn}(nn') \langle |n'|-1,0| \, e^{\frac{i}{\sqrt{2\phi}}
    (\overline{\gamma}a^\dagger+ {\gamma}a)}||n|-1,0 \rangle\\ && - \langle |n'|,0|
    e^{\frac{i}{\sqrt{2\phi}}(\overline{\gamma}a^\dagger + {\gamma}a)} ||n| ,0\rangle\biggr]
\end{eqnarray*}
and Eq.~\eqref{Eqn:H^q_{n'n}-definition} together with the form factor $F^{(1)}_{n'n}(2\pi\mathbf{G})$ in Eq.~\eqref{FCmatrixelement},
we find
\begin{widetext}
\begin{equation}
     \langle \psi_{n',0} |\sigma_z 
     e^{-2\pi i \mathbf{G}  \cdot \mathbf{r} } 
     e^{ i\mathbf{R} \cdot \mathbf{Q} }| \psi_{n,0} \rangle = 
     e^{ -\frac{(2\pi)^2}{4\phi}\overline{G}G-\frac{\phi}{4}\overline{R}R+i\pi \overline{G}R}    F^{(1)}_{n'n}(2\pi\mathbf{G}). 
\end{equation}
We next evaluate the term
\begin{equation}
   \langle \psi_{n',0} |\sigma_z   
   e^{ -2\pi i \mathbf{G} \cdot \mathbf{r} } 
   e^{ -i\mathbf{R}' \cdot \mathbf{Q} } 
   e^{ i\mathbf{R} \cdot \mathbf{Q} } |\psi_{n,0} \rangle = 
   e^{-\frac{1}{2} [ \mathbf{R}' \cdot \mathbf{Q}, \mathbf{R} \cdot \mathbf{Q} ]} \langle \psi_{n',0} | 
   \sigma_z e^{-2\pi i \mathbf{G} \cdot \mathbf{r} } 
   e^{i ( \mathbf{R}-\mathbf{R}' ) \cdot \mathbf{Q} } |\psi_{n,0} \rangle,
\end{equation}
which appears in the matrix elements in Eq.~\eqref{Eqn:Ber-matrix-elt-01}. 
Using the above relations, they are expressed in the form
\begin{eqnarray*}
       \langle \mathbf{k},n'  |  \sigma_z e^{-2\pi i \mathbf{G} \cdot \mathbf{r} } |\mathbf{k},n \rangle & = &
     \frac{1}{{\mathcal{N}(\mathbf{k})}}  
        \sum_{ \mathbf{R},\mathbf{R}'} 
        e^{-i \mathbf{k} \cdot ( \mathbf{R} -\mathbf{R}' )  
        -i\frac{\phi}{2} R'_1 R'_2+i \frac{\phi}{2} R_1 R_2
        - i\frac{\phi}{2}(R_1' R_2 -R_2' R_1)} \times \\ &\times& e^{-\frac{(2\pi)^2}{4\phi}\overline{G}G-\frac{\phi}{4}(\overline{R}-\overline{R}')(R-R')+i\pi \overline{G}(R-R')}         F^{(1)}_{n'n}(2\pi\mathbf{G}) .
\end{eqnarray*}
\end{widetext}
Performing the summations over $\mathbf{R}$ and $\mathbf{R}'$, and defining the matrix
\begin{equation}
{\cal M} = \frac{i\phi}{4\pi}  \left(\begin{array}{cc}
           1 & i\\ i & 1\end{array} \right),
\end{equation}
the matrix elements \eqref{Eqn:Ber-matrix-elt-01} follow as
\begin{eqnarray}\nonumber
 &&  \langle \mathbf{k},n'  |\sigma_z e^{-2\pi i \mathbf{G} \cdot \mathbf{r} }| \mathbf{k},n \rangle = 
   \frac{(2\pi)^2 \delta(\mathbf{0})}{{\mathcal{N}(\mathbf{k})}}  e^{-\frac{(2\pi)^2}{4\phi}\overline{G}G} \times \\ &&\quad\times\; \vartheta \bigg( \frac{(k_1-\pi \overline{G},k_2-i\pi \overline{G})}{2\pi} \bigg|{\cal M}\bigg) F^{(1)}_{n'n}(2\pi\mathbf{G}), \label{finalalmost}
\end{eqnarray}
where $\vartheta (\mathbf{z}|{\cal B})$ is the Riemann Theta function for a two-dimensional vector $\mathbf{z}=(z_1,z_2)$
and a symmetric $2\times 2$ matrix ${\cal B}$ \cite{NIST:DLMF,Mumford2007}. 

Putting $\mathbf{G} = 0$ and substituting $\sigma_z$ with the identity, we obtain
\begin{equation}
     \langle \mathbf{k},n'  |\mathbf{k},n \rangle =
     \frac{(2\pi)^2 \delta(\mathbf{0})}{{\mathcal{N}(\mathbf{k})}} \vartheta 
     \bigg( \frac{(k_1,k_2)}{2\pi} \bigg| {\cal M} \bigg) \; 
     F^{(2)}_{n'n}(\mathbf{0}), 
\end{equation}   
where we define the function
\begin{equation}
    F^{(2)}_{n'n}(2\pi\mathbf{G}) =  \mathcal{N}_n \mathcal{N}_{n'} \left[\mathrm{sgn}(n'n) 
    \mathcal{H}_{|n'|-1,|n|-1} ^{2\pi\mathbf{G}}+ \mathcal{H}_{|n'|,|n|} ^{2\pi\mathbf{G}} \right]. \nonumber
\end{equation}
With $\mathcal{H}_{|n'|,|n|}^{\mathbf{0}} = \delta_{|n'|,|n|}$, we get $F^{(2)}_{nn}(\mathbf{0}) = 1$, 
and using the orthonormality condition ~\eqref{orthonormcond},
%$\langle \mathbf{k}',n'  |\mathbf{k},n \rangle = (2\pi)^2 \delta_{n'n} \delta(\mathbf{k} - \mathbf{k}')$,  
the normalization constant follows as~\cite{Bernevig2022}
\begin{equation}
{\mathcal{N}(\mathbf{k})} = \vartheta \bigg( \frac{(k_1,k_2)}{2\pi} \bigg| {\cal M} \bigg) . 
\label{Eqn:norm-const-magnetic-bloch-state} 
\end{equation}

Finally, from Eq.~\eqref{finalalmost}, by exploiting properties of the Riemann Theta function, 
\begin{equation}
     \frac{\vartheta \left( \frac{(k_1-\pi \overline{G},k_2-i\pi \overline{G})}{2\pi} \bigg| 
     {\cal M} \right)}{\vartheta \bigg( \frac{(k_1,k_2)}{2\pi} \bigg|{\cal M} \bigg)} 
     = e^{\frac{\pi}{2} G\overline{G}+i\pi G_1 G_2 -i(k_1G_2 - k_2G_1)},
\end{equation}
we arrive at Eq.~\eqref{FCmatrixelement}.\\

\section{On Weiss oscillations}
\label{AppD}

\begin{figure}[H]
    \centering
    \includegraphics[width=7cm]{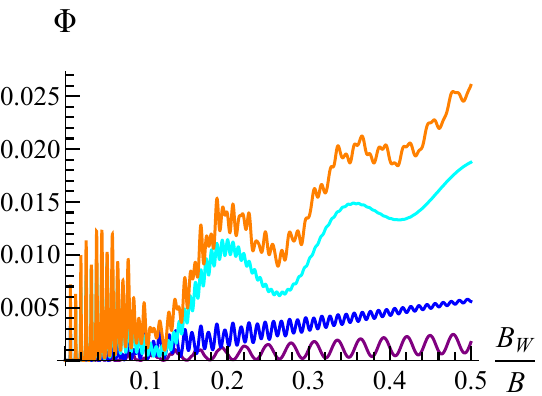}
    \caption{Dimensionless longitudinal conductivity $\Phi=\sigma_{yy}/\sigma_0$ vs inverse magnetic $1/B$, see Eq.~\eqref{App:dimcond}, 
    for the mass superlattice~\eqref{periodicprofile}. 
    We use $T = 6$\,K, $n_e = 1.5\times 10^{11}$\,cm$^{-2}$,  and
    $d = 5\bar \ell_B \simeq 128$\,nm. The corresponding Fermi energy is $90$\,meV. 
    $B$ is expressed in units of $B_W= \frac{\hbar}{e d^2}$. 
    Purple, blue, and cyan curves display the separate contributions of the
    Fourier components $m_j$ with $j = 1, 3, 5$, respectively. 
    The orange curve is obtained by retaining the first $15$ Fourier coefficients $m_j$ in Eq.~\eqref{App:dimcond}, 
    sufficient to capture all significant terms.}  
    \label{figApp2}
\end{figure}

We here provide additional details on Sec.~\ref{sec4b}. 
Using the wave functions \eqref{wf} and $X = (x-x_c)/\ell_B$, the first-order correction to the energy due to a periodic mass term $m(x)$ is given by
\begin{eqnarray} \label{firstorder}
        \Delta E_{n,k_y} & = & \mathcal{N}_n^2\int_{-\infty}^{+\infty} dX \;  m(\ell_B X+x_c)\times \nonumber\\ 
         &\times& \left [ (1-\delta_{n,0}) \Phi^2_{|n|-1} (X) -\Phi^2_{|n|} (X)  \right].
\end{eqnarray}
Expanding $m(x)$, which is assumed to be even in $x$, as a Fourier series with coefficients $m_j$, see Eq.~\eqref{cosinemass}, 
we then arrive at  Eq.~\eqref{Eqn:Delta-E-weiss-osc.} for $\Delta E_n(k_y)$ and Eq.~\eqref{velocity1} for the velocity $v_y^{n,k_y}$.
The longitudinal conductivity $\sigma_{yy}$ then follows from Eq.~\eqref{diffcond}, where the sum over $\zeta$  includes a summation over
the Landau level index $n\in\mathbb Z$ and an integration over $k_y$.
Plugging Eq.~\eqref{velocity1} into Eq.~\eqref{diffcond}, we obtain
\begin{eqnarray}
    \sigma_{yy} &=& \frac{e^2 \beta  \tau}{2\pi L_x } \left( \frac{2ud}{h}  \right)^2  
    \sum_{n = -\infty}^{+\infty} \frac{\mathcal{N}_n^4 }{4\cosh^2[\beta(E_n-\mu)/2]} \\ &\times&\nonumber
          \sum_{j,l}  P_j^n (u) P_l^n (u)\int_0^{L_x/\ell_B ^2} dk_y \sin(jk_yd_y) \, \sin(lk_yd_y),
\end{eqnarray}
with the auxiliary functions
\begin{equation}
P_j^n (u) = jm_j e^{-\frac{1}{2}j^2u} \left[ (1-\delta_{n,0}) L_{|n|-1}(j^2u)-L_{|n|}(j^2u)  \right].
\label{Eqn:Aj-form1}
\end{equation}
Next, we perform the $k_y$ integration using the identity
\begin{equation}
     \int_0^{L_x/\ell_B^2} dk_y \sin(jk_yd_y) \, \sin (j'k_yd_y) =\frac{L_x}{2\ell_B ^2}\delta_{j,j'},
\end{equation}
which holds for $L_x\gg \ell_B^2/d_y$.
We thus arrive at 
\begin{equation}
    \sigma_{yy} = \frac{ e^2 u\beta\tau}{2\pi \hbar^2 }  \sum_{n = -\infty}^{+\infty} \frac{\mathcal{N}_n^4 }{4\cosh^2(\frac{\beta}{2}(E_n-\mu))} 
    \sum_{j=1}^{\infty}  [ P_j^n (u)]^2.
    \label{App:condformula}
\end{equation}

For the step-like periodic mass profile~\eqref{periodicprofile}, the expansion coefficients $m_j$  follow from Eq.~\eqref{Eqn:Fourier-coefficients-in-2D} with $m_j = 2A_{\left( \frac{j}{d},0\right)}$.
We then obtain the dimensionless magnetoconductivity $\Phi=\sigma_{yy}/\sigma_0$, see Eq.~\eqref{sigmayy}, in the form
\begin{eqnarray}\nonumber
    \Phi &=&  \sum_{n = -\infty}^{+\infty} \frac{\mathcal{N}_n^4 u }{4\cosh^2[\beta(E_n-\mu)/2]}  
     \sum_{j=1}^{\infty} e^{-j^2u}  \sin^2\left(\frac{\pi j}{2}\right) \times \\ &\times&
    \left[ (1-\delta_{n,0}) L_{|n|-1} (j^2u)-L_{|n|} (j^2u)  \right]^2,
    \label{App:dimcond}
\end{eqnarray}
where each Fourier component $m_j$ gives an oscillatory contribution with argument $j^2 u$ 
and weight $\propto j^2 e^{-j^2 u} m_j^2$. 
The individual contributions of the first few harmonics are illustrated in Fig.~\ref{figApp2}.
The superposition of different $j$ then produces aperiodic beating patterns in $\Phi(1/B)$ as observed in Fig.~\ref{figApp2}. We note that 
the Fourier terms $m_j$ with small $j$ dominate the conductivity because of the envelope factor $e^{-j^2 u}$.

\bibliography{biblio}

\end{document}